\documentclass[12pt]{article}

\pdfoutput=1
\usepackage{amssymb,dsfont}
 \usepackage{amsfonts}
\usepackage{amssymb}
\usepackage{amsmath}
\usepackage{graphicx,psfrag,epsf}
\usepackage{enumerate}
\usepackage{natbib}
\usepackage{subcaption}
\usepackage{bbm,rotating}

\newcommand{\blind}{0}

\newcommand{\bfX}{{\bf X}}

\newcommand{\bfC}{{\bf C}}

\newcommand{\bfA}{{\bf A}}
\newcommand{\bfW}{{\bf W}}

\newcommand{\bfB}{{\bf B}}

\newcommand{\bfK}{{\bf K}}

\newcommand{\bfY}{{\bf Y}}
\newcommand{\bfz}{{\bf z}}

\newcommand{\bfI}{{\bf \mathbb{I}}}
\newcommand{\bfone}{{\mathds{1}}}

\pdfoutput=1
\begin{document}

\newcommand{\SPACEBIG}{1.45}
\newcommand{\SPACESMALL}{1}

\def\spacingset#1{\renewcommand{\baselinestretch}%
{#1}\small\normalsize} \spacingset{1}


\if0\blind
{
  \title{\bf Tests of independence and Beyond}
  \author{Julie Josse
	\hspace{.2cm}\\
    Applied Mathematics Department, Agrocampus Ouest, Rennes, France\\
    and \\
    Susan Holmes\\
    Department of Statistics, Stanford University, California, USA}
  \maketitle
} \fi

\if1\blind
{
  \bigskip
  \bigskip
  \bigskip
  \begin{center}
    {\LARGE\bf Tests of independence and Beyond
\end{center}
  \medskip
} \fi
\pdfoutput=1
\bigskip
\begin{abstract}
Simple correlation coefficients between two variables 
have been generalized to measure association between two matrices
in many ways. Coefficients such as the RV coefficient,  the distance covariance (dCov) coefficient and kernel based coefficients
have been adopted by different research communities.
Scientists use these coefficients to
test  whether two random vectors are linked.
If they are, it is important to uncover what 
patterns exist in these associations.

We discuss the topic of measures of dependence between random vectors and tests of independence and show links between different approaches.
We document some of the interesting rediscoveries and lack of interconnection between bodies of literature.
After providing definition of the coefficients and associated tests,  we present the recent improvements 
that enhance their statistical properties and ease of interpretation.
We summarize  multi-table approaches and
provide  scenarii where the indices can provide useful summaries of heterogeneous multi-block data.

We illustrate these different strategies on several examples of real data and suggest
  directions for future research.
\end{abstract}

\noindent%
\pdfoutput=1
\spacingset{\SPACEBIG}
\section{Introduction}

Today, applied statisticians
study relationships across two (or more) sets of data in many different contexts. 
A biological example  from \citet{Tayrac09}  studies 43 brain tumors of 4 different types defined by the standard world health organization (WHO) classification  (O, oligodendrogliomas; A, astrocytomas; OA, mixed oligo-astrocytomas and GBM, glioblastomas)  using data both at the transcriptome level (with expression data) and at the genome level  (with CGH data). More precisely, there are 356 continuous variables for the microarray data and 76 continuous variables for the CGH data. With  such heterogeneous data  collected  on the same samples, questions that come up include:
What are the similarities and differences between  these groups of variables? 
What is common to both groups and what is specific? 
Are two tumors that are similar at the transcriptome level also similar in terms of their genome?  To compare the information brought by each group, a first step in the analysis is to quantify the relationship between the two sets of variables using  coefficients of association and then decide if the association is significant by using a test.  Here, we  discuss the different coefficients, tests
and we will  emphasize the importance of following up a significant result with graphical representations that explore the nature of the relationships.
The analysis of the tumor data is detailed in Section \ref{sec:microarray}.

Studying and assessing the relationship between two sets of data can be traced back to the work of \citet{David1962, Barton1962, Knox} and \cite{David}. 
Their aim was to study  space-time association to detect epidemics in diseases such as leukemia. 
To do so, they computed two distance matrices, one measuring the differences in time 
between disease occurrences, the other measuring 
the spatial distance between events. 
Then, they thresholded the largest distances, built a graph from each matrix and computed as a measure of relationship the number of edges in the intersection of the two graphs. A high association indicated a high chance
of an occurrence of an epidemic. 
Asymptotic tests were suggested to evaluate the evidence for an association.
Although not referring to graphs, \citet{Mantel} directly computed the correlation coefficient between the two lower triangular parts of the
distance matrices and used a permutation test
to detect significance. 
His name is now associated to 
this popular method of randomized testing between two
distance matrices. 

Many different coefficients and tests have been published as measures of
association between two data tables, popular ones are 
 the RV coefficient \citep{Escoufier70}, the Procrustes coefficient \citep{Gow71}  and more recently the dCov coefficient \citep{Szekely07}. Two points are striking when investigating this topic. First,  the citation record of  papers covering the subject shows that different disciplines have adopted
different types of coefficients with strong within discipline preferences. 
If we look at the list of the 7,000
papers citing \citet{Mantel}, ranked according to citations, more than half of the books and references are in the ecological and genetic disciplines, with other areas that use
spatial statistics intensively  well represented. 
 Of the 370 papers citing the original RV papers \citet{Escoufier70,Escoufier73}, 
almost half are methodological papers which do not have a particular field of application, of the others 
40\% come from ecology, almost 30\% come from food science and sensory analyses,
whereas 20\% originate from neuroscience, other well represented disciplinary areas are chemometrics,
shape analyses and genomics. 
The Procrustes coefficient \citep{Gow71}, is cited more than 1000 times and is very popular in ecology, morphology and neuroscience. 
Although recent, about a hundred  papers cite the dCov coefficient \citep{Szekely07}, most of which are theoretical. 
The second main point is that there are rediscoveries and lack of interconnection between the bodies of literature.
For instance,  \citet{Szekely07} introduced the distance covariance (dCov) coefficient which has the property of being equal to zero if and only if the random vectors are independent. 
This coefficient aroused the interest of the statistical community and gave rise to much research on this topic. 
\citet{Sejdinovic13} made the link between the Hilbert-Schmidt Independence Criterion (HSIC),  a kernel based measure of independence developed in the machine learning community  \citep{Gretton05}, and the dCov coefficient. This literature on the dCov coefficient and on the kernel based coefficients, however, seems to have overlooked the literature on the RV coefficient despite many common features which will be illustrated in this paper. 
The RV coefficient can be seen as an early instance of a natural generalization of the notion of correlation to groups of variables. 

\citet{Laz88} and \citet{ramsay84} discussed more than 10 other coefficients, including
 the early canonical correlation coefficient \citep{Hotelling36}.
 \citet{Beran20071805},  \citet{Koja}  and \citet{Quessy20103058} defined coefficients using an empirical 
 process point of view. 
Some coefficients have been completely forgotten, the coefficients that thrive are  the ones 
 implemented in mainstream
software. 

In this paper, we focus on three classes of coefficient in current use. First, we consider linear relationships 
that can be detected with  the RV coefficient presented in Section \ref{sec_RV}. After giving some of its properties, 
we present two modified versions of the RV coefficient  proposed to correct the potential sources of bias.  We conclude this Section by presenting two other coefficients based on linear relationships, the Procrustes coefficient \citep{Gow71} and the $L_g$ coefficient \citep{Escofier1994121, Pages14}. 
Section \ref{sec:dcov} focuses on the detection of
 non-linear relationships using the dCov coefficient.  Covering  the same
 topics (asymptotic tests, permutation tests, modified coefficients) 
 for both the 
 RV and the dCov coefficients allows us to highlight their similarities.
We show by a small simulation a  comparison of these coefficients.
The RV coefficient and the dCov coefficient rely on Euclidean distances, squared Euclidean for the former and Euclidean for the latter. We discuss  in Section \ref{sec:beyond} coefficients that can be based on other distances or dissimilarities such as the Mantel coefficient  \citep{Mantel}, a graph based measure defined by \cite{Frie83} and  the HSIC coefficient \citep{Gretton05}.
Finally, in Section \ref{realdata}, we illustrate the  practical use of these coefficients on real data sets coming from sensory analysis, genetics, morphology and chemometry. We highlight graphical methods for the exploration of the relationships and make some suggestions for some of the coefficients lacking these
follow-up tools.

\section{The RV coefficient} \label{sec_RV}

\subsection{Definition} \label{defRV} 

Consider two random vectors $X$ in $\mathbb{R}^p$ and $Y$ in $\mathbb{R}^q$. Our aim is to study and test the association between these two vectors. 
We represent  $n$ independent realizations of the random vectors
by matrices $\bfX_{n \times p}$ and $\bfY_{n \times q}$, which we assume column-centered.

The rationale underlying the RV coefficient is to consider that two sets of variables are correlated if the relative position of the observations in one set is similar to the relative position of the samples in the other set. The matrices representing the relative positions of the observations are the cross-product matrices: $\bfW_{\bfX}=\bfX\bfX' $ and $\bfW_{\bfY}=\bfY\bfY'$. They are of size $n \times n$ and  can be compared directly. To measure their proximity, the Hilbert-Schmidt inner product between matrices is computed:
\begin{eqnarray}
<\bfW_{\bfX},\bfW_{\bfY}>=\mbox{tr}(\bfX\bfX' \bfY\bfY')= \sum_{l=1}^{p} \sum_{m=1}^{q} \mbox{cov}^2(\bfX_{.l}, \bfY_{.m}),
\label{num_RV}
\end{eqnarray}
with $\mbox{cov}$ the sample covariance coefficient and $\bfX_{.l}$ the column $l$ of matrix $\bfX$ and $\bfY_{.m}$ the column $m$ of matrix $\bfY$. Since the two matrices $\bfW_{\bfX}$ and $\bfW_{\bfY}$ may have different norms,  a correlation coefficient, called the RV coefficient by \citet{Escoufier73}, is computed by normalizing by the matrix norms:

\begin{eqnarray}
\mbox{RV}(\bfX,\bfY)&=& \frac{<\bfW_{\bfX},\bfW_{\bfY}>}{ \parallel \bfW_{\bfX}\parallel
\parallel \bfW_{\bfY} \parallel } = \frac{\mbox{tr}(\bfX\bfX' \bfY\bfY')}{\sqrt{\mbox{tr}(\bfX\bfX')^2
\mbox{tr}(\bfY\bfY')^2  }}. 
\label{RV}
\end{eqnarray}
This computes the cosine of the angle between the two vectors in $\mathbb{R}^{n \times n}$ representing the cross-product matrices. 
It may be convenient to write the RV coefficient in a different way to  understand its properties, for instance using the covariance matrices: 
$\mbox{RV}(\bfX,\bfY)= \frac{\mbox{tr}(S_{\bfX\bfY}S_{\bfY\bfX})}{\sqrt{\mbox{tr}(S_{\bfX\bfX}^2)
\mbox{tr}(S_{\bfY\bfY}^2)}}$, with $S_{\bfX\bfY}=\frac{1}{n-1}\bfX'\bfY$ being the empirical covariance matrix between $\bfX$ and $\bfY$.
It is also possible to express the coefficient using distance matrices. More precisely, let $\boldsymbol{\Delta}_{n \times n}$ be the matrix where element $d_{ij}$ represents the Euclidean distance between the observations $i$ and $j$, $d_{i.}$ and $d_{.j}$ being the mean of the row $i$ and the mean of column $j$ and $d_{..}$ being the global mean of the distance matrix. Using the formulae relating the cross-product and the Euclidean distance between two observations \citep{Schoenberg35, Gower66},
$W_{ij}=-\frac{1}{2}(d_{ij}^2-d_{i.}^2-d_{.j}^2+d_{..}^2)$, the RV coefficient \eqref{RV} can be written as: 

\begin{eqnarray}
\mbox{RV}(\bfX,\bfY)= \frac{<\bfC\boldsymbol{\Delta}_{\bfX}^2\bfC,\bfC\boldsymbol{\Delta}_{\bfY}^2\bfC>}{ \parallel \bfC\boldsymbol{\Delta}_{\bfX}^2\bfC\parallel
\parallel \bfC\boldsymbol{\Delta}_{\bfY}^2\bfC \parallel },
\label{RVdist}
\end{eqnarray}
 with $\bfC=\boldsymbol{\bfI_n} - \frac{\boldsymbol{\bfone_n}\boldsymbol{\bfone_n}'}{n}$, $\boldsymbol{\bfI_n}$ the identity matrix of order $n$ and  $\bfone_n$ a vector of ones of size $n$. The numerator of \eqref{RVdist} is the inner product between the double centered (by rows and by columns) squared Euclidean distance matrices. This latter expression \eqref{RVdist} will be important for the sequel of the paper since it enables an easy comparison with other coefficients.

The population counterpart of the RV coefficient is called the vector correlation coefficient $\rho_V$ and is often expressed with the following form $\rho_V(X,Y)= \frac{\mbox{tr}(\Sigma_{XY}\Sigma_{YX})}{\sqrt{\mbox{tr}(\Sigma_{XX}^2)
\mbox{tr}(\Sigma_{YY}^2)}}$, with $\Sigma_{XY}$ the population covariance matrix between $X$ and $Y$. \\

Some of the properties of the RV coefficient are:
\begin{itemize}
\item Statistical consistency when  $n \rightarrow \infty$.
\item for $p=q=1$, $\mbox{RV}= r^2$ the square of the standard correlation coefficient
\item $ 0 \leq \mbox{RV}(\bfX,\bfY)\leq 1 $
\item $ \mbox{RV}(\bfX,\bfY)=0$ if and only if $\bfX'\bfY=0$: all the variables of one group are orthogonal to all the variables in the other group.
\item $\mbox{RV}(\bfX,a\bfB \bfX +\bold{c})=1$, with $\bfB$ an orthogonal matrix, $a$ a constant and $\bold{c}$ a constant vector of size $n$. The RV is invariant by shift, rotation, and overall scaling.  
\end{itemize}
Remark:\\
\begin{enumerate}
\item If the column-variables of both matrices $\bfX$ and $\bfY$ are standardized to have unit variances, the numerator of the RV coefficient \eqref{num_RV} is equal to the sum of the squared correlations between the variables of the first group and the variables of the second group. It is thus crucial to consider what ``pre-processing" has been undertaken on the data when analyzing the
coefficient. 

\item The RV can be seen as an ``unifying tool" that encompasses many methods derived by maximizing the association coefficients under specific constraints. \citet{Rob76} showed for instance that the PCA of $\bfX$ seen as  maximizing RV$(\bfX,\bfY=\bfX\bfA)$ with $\bfA$ being an $n \times k$ matrix under the constraints that $\bfY'\bfY$ is diagonal. Discriminant analysis, canonical analysis as well as multivariate regression can also be derived in the same way.
\end{enumerate}

\subsection{Tests}

As with the ordinary correlation coefficient, a high value of the RV coefficient does not 
necessarily mean there is  a significant relationship between the two sets of measurements. We will show in Section \ref{permut} that the  RV coefficient depends on both the sample size and on the covariance structure of each matrix; hence the need for a valid
inferential procedure for testing the significance of the association. 
One usually sets up the hypothesis test by taking
\begin{displaymath}
    \left\{ \begin{array}{lcl}
        H_{0} && \rho V=0, \textrm{there is no association between the two sets}
        \\
        H_{1} && \rho V > 0, \textrm{there is an association between the two sets}
    \end{array} \right.
\end{displaymath}
The test evaluates the strength of any linear relationship between the two sets. 
The fact that $\rho V=0$ (which corresponds to the population covariance matrix $\Sigma_{XY}=0$) does not necessarily imply independence between $X$ and $Y$ (except when they are multivariate normal), only the absence of a linear relationship between them.

\subsubsection{Asymptotic tests \label{sec:asympt:rv}}
Under the null hypothesis, the asymptotic distribution of the $n$RV  is available when the joint distribution of the random variables is multivariate normal or when it belongs to the class of elliptical distributions \citep{cle89}.  Precisely, \citet{Rob85} showed that under those assumptions, $n$RV converges to:
\begin{eqnarray}
\frac{1+k}{\mbox{tr}(\Sigma^2_{XX})\mbox{tr}(\Sigma^2_{YY})} \sum_{l=1}^{p} \sum_{m=1}^{q} \lambda_l \gamma_m Z_{lm}^2,
\label{test_rv_asymp}
\end{eqnarray}
where:\\
$k$ is the kurtosis parameter of the elliptical distribution,\\
$\lambda_1 \geq \lambda_2 \geq...\geq \lambda_p$ are the eigenvalues of the covariance matrix $\Sigma_{XX}$,\\
$\gamma_1  \geq \gamma_2  \geq... \geq \gamma_q$ are the eigenvalues of the covariance matrix $\Sigma_{YY}$,\\
and $Z_{lm}$ are i.i.d  $\mathcal{N} (0,1)$ random variables.\\
To eliminate the need for any distributional hypotheses, \cite{cle95} suggested a test based on the rank. However, \cite{Josse08} showed that these tests only provide accurate type I errors for large sample sizes ($n>300$). An alternative is to use permutation tests.
\subsubsection{Permutation tests \label{permut}}
Permutation tests were used
to ascertain a links between two sets of variables in the earliest
instance of multi-table association testing.
Repeated  permutation of the 
rows of one matrix and computation of the  statistic
such as the RV coefficient provides the null
 distribution of no association.
There are $n!$ possible permutations
to consider and the $p$-value is the proportion of the values that are greater or equal to the observed coefficient. 
 
Note that care must be taken in the implementation as this is not equivalent to a complete 
permutation test of the vectorized cross-product matrices for which the exhaustive distribution is much
larger $(n(n-1)/2!)$.

 Computing the exact permutation distribution is computationally costly when $n>15$. Consequently, the permutation distribution is usually approximated by Monte Carlo, although a moment matching approach is also possible. 
 The latter consists of approximating the permutation distribution by a continuous distribution without doing any permutation and using the analytical moment of the exact permutation distribution under the null.
\cite{Kaz} defined the first moments of the quantity \eqref{num_RV} under the null which yields the moments of the RV coefficient. 
The expectation is:
\begin{eqnarray}
{\mathbb E}_{H_0}(\mbox{RV})=\frac{\sqrt{\beta_x \times
\beta_y}}{n-1} ~~ \mbox{with} ~~\beta_x &=& \frac{(\mbox{tr}(\bfX'\bfX))^2}{\mbox{tr}((\bfX'\bfX)^2)}= \frac{(\sum
\lambda_i)^2}{\sum \lambda^{2}_{i}}
\label{expectation}
\end{eqnarray}
and $\beta_y$ is defined similarly.
Equation \eqref{expectation} provides insight into the expected behavior of the RV coefficient
with $\beta_x$  providing a measure of the complexity of the matrix. The coefficient varies between $1$ when all the variables are perfectly correlated and $p$ when all the variables are orthogonal. Thus, equation \eqref{expectation} shows that under the null, the RV coefficient takes high values when the sample size is small (as with the simple correlation coefficient) and when the data matrices $\bfX$ and $\bfY$ are very multi-dimensional. The expression of the variance and the skewness are detailed in \cite{Josse08}. With the first three moments, \cite{Josse08} compared different moment based methods such as the Edgeworth expansions or the Pearson family and  pointed out the quality of the Pearson type III approximation for permutation distributions. 
The RV based tests are implemented in the R \citep{CRAN} packages {\tt ade4} \citep{dray07} as {\tt RV.rtest} and as {\tt coeffRV} in {\tt FactoMineR} \citep{Husson13}. The former uses 
Monte Carlo generation of the permutations whereas  the latter uses a Pearson type III approximation.

\subsection{Modified coefficients \label{sec:rv:biais}}
In practice,  the statistical significance of test
is not informative enough and 
 one may also want to quantify and decompose the association.
 
Equation \eqref{expectation} shows the  RV value is not informative on its own
as it  depends  on the sample size. As underlined by \cite{Smilde09} 
and independently by \citet{Kaz} and  \citet{Josse08}  even under the null, the values of the RV coefficient can be very high.
 For this reason modified versions of the coefficient have been suggested. 
 
  By computing expectations under the null of the coefficient for two independent normal
  random matrices $\bfX$ and $\bfY$ using random matrix theory,
   \cite{Smilde09}  showed that the problem can be traced back to the diagonal elements of the matrices $\bfX\bfX'$ and $\bfY\bfY'$. Thus, they proposed a new coefficient, the modified RV, by removing those elements: 
\begin{eqnarray}
\mbox{RV}_{\rm mod}(\bfX,\bfY) = \frac{\mbox{tr}((\bfX\bfX'- \mbox{diag}(\bfX\bfX')) (\bfY\bfY'- \mbox{diag}(\bfY\bfY')))}{\sqrt{\mbox{tr}(\bfX'\bfX - \mbox{diag}(\bfX\bfX'))^2
\mbox{tr}(\bfY'\bfY - \mbox{diag}(\bfX\bfX'))^2 }}. 
\label{RVmod}
\end{eqnarray}
This new coefficient can take on negative values. They showed in a simulation study that their coefficient has the expected behavior, meaning that even in high dimensional setting ($n=20$ and $p=q=100$), the values of the  $\mbox{RV}_{\rm mod}$ are around 0 under the null. In addition, for a fixed value of $n$, they simulated two matrices uncorrelated to each other and slowly increased the correlation between the two groups. They showed that the $\mbox{RV}_{\rm mod}$ varies between 0 and 1 whereas the RV varies between 0.85 to 0.99. Thus, they argued that the modified coefficient is  easier to interpret. \\

We can make a connection between the debiased coefficient and copulas \citep{Nelsen06} which aim at removing the marginal effects to focus on the structure of dependence. 
Note that  \citet{Green88, Green94} also tried to remove the diagonal terms of the cross-product matrix in the method joint correspondence analysis (JCA) to fit only the non-diagonal part of the Burt matrix (the matrix that cross tabulates all the categorical variables). \\

\citet{Mayer11} extended \cite{Smilde09}'s work by highlighting the fact that the $\mbox{RV}_{\rm mod}$ \eqref{RVmod} is still biased under the null. The rationale of \citet{Mayer11}'s approach is to replace the simple correlation coefficient $r^2$ in the expression of the RV coefficient (which can be seen in equation \eqref{num_RV} when the variables are standardized) by an adjusted coefficient. They only considered the case of standardized variables.
More precisely, they defined the adjusted RV as:
\begin{eqnarray*}
\mbox{RV}_{\rm adj}= \frac{\sum_{l=1}^p\sum_{m=1}^q r^2_{\rm adj}(\bfX_{.l},\bfY_{.m})}{\sqrt{\sum_{l,l'=1}^p r^2_{\rm adj}(\bfX_{.l},\bfX_{.l'}) \sum_{m,m'=1}^q r^2_{\rm adj}(\bfY_{.m},\bfY_{.m'})}},\\ \mbox{ with } r^2_{\rm adj}=1-\frac{n-1}{n-2}(1-r^2). \nonumber
\label{adjustRV*}
\end{eqnarray*}
A permutation test performed using this coefficient gives the same
results as that with the RV because the two statistics are equivalent,
the denominator being invariant under permutation and the numerator is monotone.
In their simulation study, they focused on the comparison between $\mbox{RV}_{\rm adj}$ and $\mbox{RV}_{\rm mod}$ by computing the mean square error (MSE) between the sample coefficients and the population coefficient ($\rho V$) and showed smaller MSE with their new coefficient. We stress this approach here, as very few
papers studying these coefficients  refer to a population coefficient. 

Both \cite{Smilde09} and \cite{Mayer11} used their coefficients on real data from biology (such as samples described by groups of genes) and emphasized the relevant interpretation from a biological perspective. In addition, \cite{Mayer11} applied a multidimensional scaling (MDS, PCoA) projection \citep{Borg05}  of  the matrix of adjusted RV coefficients between the groups of genes showing similarities between the groups.  
Such an analysis is comparable to the earlier STATIS approach where \citet{Escou87}
uses  the matrix of the RV coefficients  to compute a compromise eigenstructure on which to project each table (as illustrated in Section \ref{sec:panel}).  

The important steps of such an approach are thus, compute the coefficient, study its significance and then
visualize the relationships through a dimension reduction technique applied at the level of blocks of variables.


\subsection{Fields of application}

The RV coefficient is a standard measurement in many fields.
For instance, in sensory analysis, the same products (such as wines, yogurts or  fruit) can be described by both sensory descriptor variables (such as bitterness, sweetness or texture) and  physical-chemical measurements (such as pH, NaCl or sugars). Scientists often need ways of  comparing the sensory profile with the chemical  one \citep{Genard, Pag05}. 
Other references in sensory analysis include \citet{Schlich1996,Risvik1997,Noble2002,giacalone2013consumer,Cadena2013}.
The RV coefficient has also been successfully applied in morphology \citep{Klingenberg09,Carmelo13,Santana2013,foth2013intraspecific}, neuroscience where \citet{Malave2008} and \cite{Abdi10} used it to compute the level of association between stimuli and brain images captured using fMRI and in transcriptomics where, for instance, \citet{Culhane2003} used it to assess the similarity of expression measurements done with different technologies.

\subsection{Other linear coefficients}

\subsubsection{The Procrustes coefficient. \label{sec:proc}} 

The Procrustes coefficient \citep{Gow71} also known as the Lingoes and 
Sch\"onemann (RLS) coefficient \citep{Ling74} is defined as follows:
\begin{eqnarray}
\mbox{RLS}(\bfX,\bfY)= \frac{\mbox{tr}(\bfX\bfX' \bfY\bfY')^{1/2}}{\sqrt{\mbox{tr}(\bfX'\bfX)
\mbox{tr}(\bfY'\bfY) }}. 
\label{procu}
\end{eqnarray}
Its properties are close to those of the RV coefficient. When $p=q=1$, RLS is equal to $|r|$.  It varies between 0 and 1, being equal to 0 when $\bfX'\bfY=0$ and to 1 when one matrix is equivalent to the other
 up to an orthogonal transformation.  
\citet{Laz92} showed that $\sqrt{pq} \mbox{RLS}^2 \leq \mbox{RV}\leq \frac{1}{\sqrt{pq}} \mbox{RLS}^2$. 

To assess the significance of the RLS coefficient, a permutation test \citep{Jackson95, Neto01} is used. The coefficient and the test are implemented in the R package {\tt ade4} \citep{dray07} in the function {\tt procuste.randtest} and in the R package {\tt vegan} \citep{vegan}  in the function {\tt protest}.  Based on some simulations and real datasets, the tests based on the RV and on the Procrustes coefficients are known to give roughly similar results \citep{dray03} in terms of power. 
The use of this Procrustes version is widespread in morphometrics \citep{Rohlf1990} since the rationale of Procrustes analysis is to find the optimal translation, rotation and dilatation that superimposes configurations of points. Ecologists also use this coefficient to assess the relationship between tables \citep{Jackson95}.

\subsubsection{The $L_g$ coefficient. \label{sec:lgcoeff}}

The $L_g$ coefficient \citep{Escofier1994121} is at the core of a multi-block method named multiple factor analysis (MFA) described in \citet{Pages14}. 
At first, it is presented to assess  the relationship between one variable $\bfz_{n \times 1}$ and a group $\bfX$ as: 
\begin{eqnarray*}
L_g(\bfz,\bfX ) = <\frac{\bfW_{\bfX}}{\lambda_1}, \bfz\bfz'>=\frac{1}{\lambda_1} \sum_{l=1}^{p} \mbox{cov}^2(\bfX_{.l}, \bfz),
\end{eqnarray*}
with $\lambda_1$ the first eigenvalue of the empirical covariance matrix of $\bfX$. Thus, this coefficient varies from 0 when all the variables of $\bfX$ are uncorrelated to $\bfz$ and 1 when the first principal component of $\bfX$ coincides with $\bfz$. The coefficient for one group is $L_g(\bfX,\bfX )= \sum_{l=1}^{p} \frac{\lambda_l}{\lambda_1}= 1+ \sum_{l=2}^{p} \frac{\lambda_l}{\lambda_1}$. It can be interpreted as a measure of dimensionality with high values indicating a multi-dimensional group.  Finally, between two groups, the measure is:
\begin{eqnarray}
L_g(\bfX,\bfY ) = <\frac{\bfW_{\bfX}}{\lambda_1}, \frac{\bfW_{\bfY}}{\gamma_1}>,
\nonumber
\end{eqnarray}
with $\gamma_1$ the first eigenvalue of the empirical covariance matrix of $\bfY$. This measure is all the more important than the two groups are multi-dimensional and share dimensions which are important dimensions within each group. \citet{Pages14} provided a detail comparison between the RV coefficient and the $L_g$ one highlighting the complementary use of both coefficients.  For instance, in a situation where  $\bfX$ has two strong dimensions (two blocks of correlated variables) and $\bfY$ has the same two dimensions but in addition, it has many independent variables, the RV coefficient tends to be small whereas the Lg coefficient focuses on what is shared and takes a relatively high value.  
As  \citet{ramsay84} said  ``\textit{Matrices may be similar or dissimilar in a great many ways, and it is desirable in practice to capture some aspects of matrix relationships while ignoring others.}'' 
As in the interpretation of any statistic based on distances, it is important to understand what
similarity is the focus of the measurement, as already pointed out by \cite{Reimherr13}, the task is not easy. 
It becomes even more involved for coefficients that measure non linear relations as detailed in the next section.
\section{The dCov coefficient} \label{sec:dcov}
\cite{Szekely07} defined a  measure of dependence between random vectors: the distance covariance (dCov) coefficient that 
 had a strong impact on the statistical community \citep{newton2009}. 
The authors showed that for all random variables with finite first moments, the dCov coefficient generalizes the idea of correlation in two ways. First, this coefficient can be applied when $X$ and $Y$ are of any dimensions and not only for the simple case 
where $p=q=1$. They constructed their coefficient as a generalization
of  the simple correlation coefficient without reference to the earlier RV literature.
Second, the dCov coefficient is equal to zero, if and only if there is independence between the random vectors. Indeed, a correlation coefficient measures linear relationships and can be equal to 0 even when the variables are related. This can be seen as a major shortcoming of the correlation coefficient and of the RV coefficient.  \cite{renyi}  already pinpointed this drawback of the correlation coefficient when defining the properties that a measure of dependence should have. 

The dCov coefficient is defined as a weighted L$^2$ distance between the joint and the product of the marginal characteristic functions of the random vectors. The choice of the weights is crucial and ensures the zero-independence property. 
Note that the dCov can be seen as a special case of the general idea discussed in \citet{Romano88, Romano89} which consists in comparing the product of the empirical  marginal distributions to their joint distribution using any statistic that detects dependence. The dCov uses the characteristic functions. 
The dCov coefficient can also be written in terms of the expectations of Euclidean distances which is easier to interpret:
\begin{eqnarray}
\mathcal{V}^{2}&=&{\mathbb E}(|X-X'||Y-Y'|)+{\mathbb E}(|X-X'|){\mathbb E}(|Y-Y'|) \label{dcov}  \\&&-{\mathbb E}(|X-X'||Y-Y''|) -{\mathbb E}(|X-X''||Y-Y'| \nonumber  \\
&=&\mbox{cov}(|X-X'|,|Y-Y'|)-2\mbox{cov}(|X-X'||Y-Y''|). \label{dcovcov} 
\end{eqnarray}
with $(X', Y')$ and $(X'',Y'')$ being independent copies of $(X,Y)$  and $|X-X'|$ being the Euclidean distance (we stick to their notation). Expression \eqref{dcovcov} shows that when the covariance of the distances is equal to 0 the dCov coefficient is not necessarely equal to 0 (there is no independence).  Expression \eqref{dcov} implies a straightforward empirical estimate $\mathcal{V}^{2}_n(\bfX,\bfY)$ also known as $\mbox{dCov}_n^{2}(\bfX,\bfY)$:
\begin{eqnarray*}
\mbox{dCov}_n^{2}(\bfX,\bfY) &=& \frac{1}{n^2} \sum_{i,j=1}^n d_{ij}^{\bfX} d_{ij}^{\bfY} + d_{..}^{\bfX} d_{..}^{\bfY} - 2 \frac{1}{n} \sum_{i=1}^n d_{i.}^{\bfX} d_{i.}^{\bfY}\\
&=& \frac{1}{n^2} \sum_{i,j=1}^n (d_{ij}^{\bfX}-d_{i.}^{\bfX}-d_{.j}^{\bfX}+d_{..}^{\bfX}) (d_{ij}^{\bfY}-d_{i.}^{\bfY}-d_{.j}^{\bfY}+d_{..}^{\bfY}).
\end{eqnarray*}
Once the covariance  defined,  the corresponding correlation coefficient $\mathcal{R}$ 
is obtained by standardization. Its empirical estimate $\mbox{dCor}_n^{2}$ is thus defined as: 
\begin{eqnarray}
\mbox{dCor}_n^2(\bfX,\bfY)&=& \frac{<\bfC\boldsymbol{\Delta}_{\bfX}\bfC,\bfC\boldsymbol{\Delta}_{\bfY}\bfC>}{ \parallel \bfC\boldsymbol{\Delta}_{\bfX}\bfC\parallel
\parallel \bfC\boldsymbol{\Delta}_{\bfY}\bfC \parallel }. 
\label{dcor}
\end{eqnarray}
The only difference between this and the RV coefficient \eqref{RVdist}
is that Euclidean distances $\boldsymbol{\Delta}_{\bfX}$ and $\boldsymbol{\Delta}_{\bfY}$ are used in \eqref{dcor} instead of their squares. This difference implies that the dCor coefficient  detects non-linear relationships whereas the RV coefficient is restricted to linear ones. Indeed, when squaring distances, 
many terms cancel whereas when the distances are not squared, 
no cancellation occurs allowing more complex associations to be detected.\\

The properties of the coefficient are: 
\begin{itemize}
\item Statistical consistency  when $n \rightarrow \infty$
\item $p=q=1$ with Gaussian distribution: $\mbox{dCor}_n\leq|r|$, \\
$\mbox{dCor}^2=\frac{r \mbox{arcsin}(r)+ \sqrt{(1-r^2)}- r \mbox{arcsin}(\frac{r}{2})-\sqrt{4-r^2}+1} {1+\frac{\pi}{3}-\sqrt{3}}$
\item $0 \leq \mbox{dCor}_n(\bfX,\bfY) \leq 1$
\item $\mathcal{R}(X,Y)=0$ if and only if $X$ and $Y$ are independent 
\item $\mbox{dCor}_n(\bfX,a\bfX  \bfB +\bold{c})=1$ 
\end{itemize}
Note the similarities to some of the properties of the RV coefficient (Section \ref{defRV}). 
Now, as in Section \ref{sec_RV}, derivations of asymptotic and permutation tests and extensions to modified coefficients are provided.

\subsection{Tests \label{sec:tests:dcov}}

\subsubsection{Asymptotic test \label{sec:asympt:dcov}}
Asymptotic test is derived to evaluate the evidence of relationship between the two sets. An appealing property of the distance correlation coefficient is that the associated test assesses independence between the random vectors. \cite{Szekely07} showed that under the null hypothesis of independence, $n\mathcal{V}^{2}_n$ converges in distribution to a quadratic form: $Q= \sum_{j=1}^{\infty} \eta_j Z_j^2$, where $Z_j$ are independent standard Gaussian variables and  $\eta_j$ depend on the distribution of $(X,Y)$. Under the null, the expectation of $Q$ is equal to 1 and it tends to infinity otherwise. Thus, the null hypothesis is rejected for large values of $n\mathcal{V}^{2}_n(\bfX,\bfY)$.
One main feature of this test is that it is consistent against all dependent alternatives whereas some alternatives are ignored in the test based on the RV coefficient \eqref{test_rv_asymp}.  

\subsubsection{Permutation tests}

Permutation tests are used to assess the significance of the distance covariance coefficient in practice. The coefficient and its test are implemented in the R package {\tt energy} \citep{Szek_ener} in the function {\tt dcov.test}. Monte Carlo is used to generate a random subset of permutation. Methods using  moment based approximations could be considered for this coefficient.

\subsection{Modified coefficients}

As in \cite{Smilde09}, \cite{Szekely13b} remarked that the $\mbox{dCor}_n$ coefficient can take high values even under independence especially in high-dimensional settings. In addition, they showed that $\mbox{dCor}_n$ tends to 1 when $p$ and $q$ tend to infinity. 
Thus, they defined a corrected coefficient dCor*($\bfX$,$\bfY$) to make the interpretation easier. The rationale is to remove the bias under the null \citep{Szekely13}. 
The dCor* coefficient can take negative values. Its distribution under the null in the modern setting where $p$ and $q$ tend to infinity has been derived and can be used to perform a test. This coefficient and the test are implemented in the function {\tt dcor.ttest}. 

\subsection{Generalization  \label{sec:dcov_alpha}}

\citet{Szekely07} showed that the theory still holds when the Euclidean distance $d_{ij}$ is replaced by $d_{ij}^{\alpha}$ with $0 \leq \alpha < 2$. This means that a whole set of coefficients can be derived and that the tests will still be consistent against all alternatives. 
 Consequently, dCov with exponent ${\alpha}$ generalizes the RV which is the same as dCor$^2$ with $\alpha=2$.

\subsection{Simulations \label{simu}}

To assess the performance of the dCov coefficient and the RV coefficient, we reproduce similar simulations to those in \cite{Szekely07} adding the comparison to the RV coefficient.

\begin{figure}[!ht]
\includegraphics[width=0.45\textwidth]{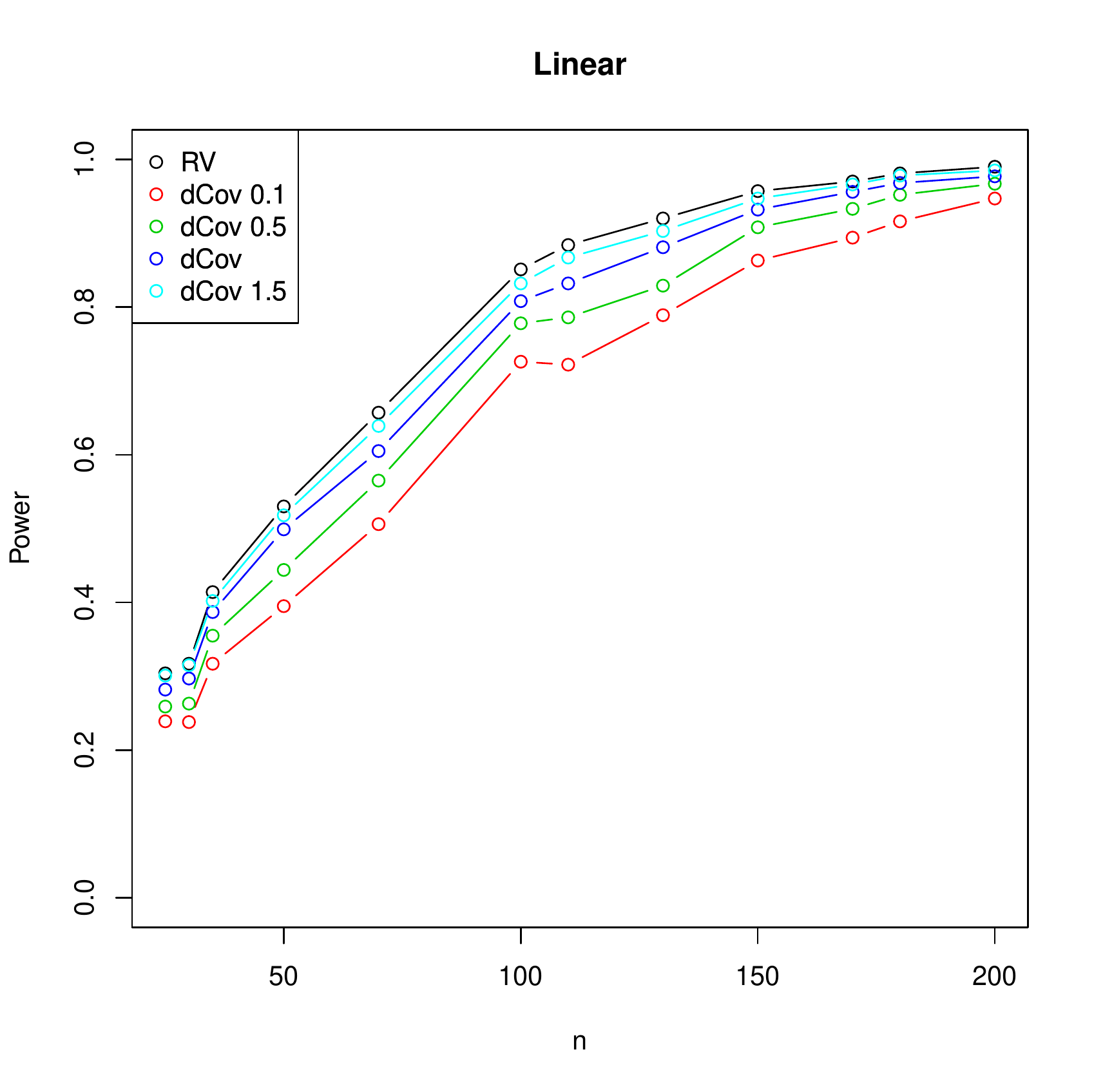}
\includegraphics[width=0.45\textwidth]{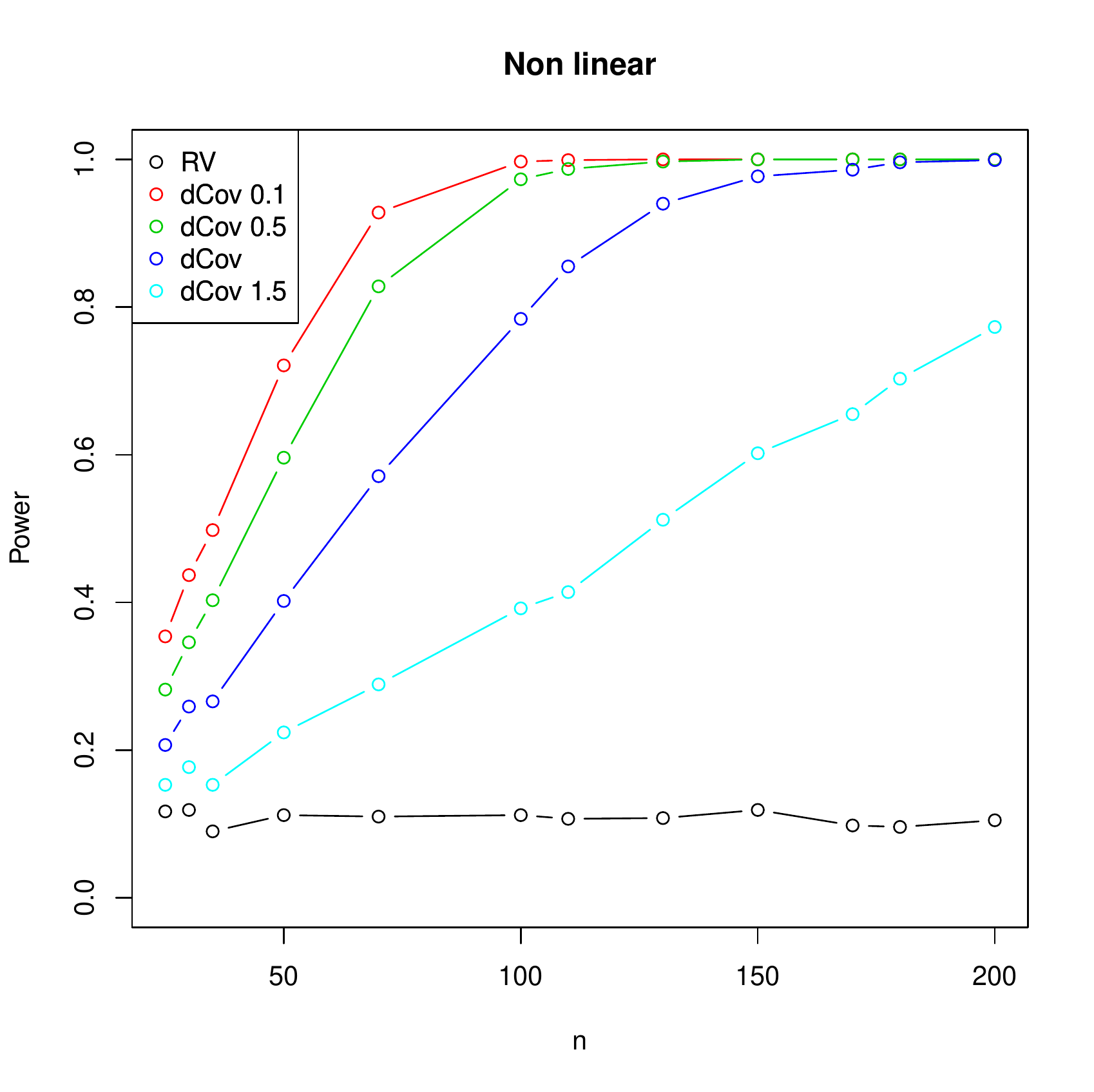} 
\caption{Power of the RV test and dCov test. Left: linear case. Right: non-linear case. The dCov test is performed using different exponents $\alpha$ (0.1, 1, 1.5, 2) on the Euclidean distances.}
\label{puissance-dcov}
\end{figure}

  
First, matrices $\bfX_{n \times 5}$ and $\bfY_{n \times 5}$ were generated from a multivariate Gaussian distribution with a within-matrix covariance structure equal to the identity matrix and the covariances between all the variables of $\bfX$ and $\bfY$ equals to $0.1$. We generated 1000 draws and computed the RV test (using the Pearson approximation) as well as the dCov test (using 500 permutations) for each draw. Figure \ref{puissance-dcov}, on the left, shows the power of the tests for different sample sizes $n$ demonstrating the similar behavior of the RV (black curve) and dCov (dark blue curve) tests with a small advantage for the RV test. 
We also added the tests using different exponents $\alpha=(0.1, 0.5,1.5)$ on the Euclidean distances which lead to  different performances in terms of power. 
 
Then, another data structure was simulated by generating the matrix $\bfY$ such that $\bfY_{ml}=log(\bfX_{ml}^2)$ for $m,l=1,...,5$ and the same procedure was applied. Results are displayed in Figure \ref{puissance-dcov} on the right. 
As expected, the  dCov tests  are more powerful than the RV test in this non-linear setting. 

These results show that  the dCov  detects  linear relationships and has the advantage of detecting other associations, so is a considerable improvement on the
 RV  and other `linear' coefficients. However, it may still be worth using the RV coefficient for two reasons. 
First, with a significant dCov, it is impossible to know the pattern of association: are there only linear relationships between variables? only non-linear relationships or both kinds? Consequently, from a practical point of view, performing both the dCov and RV tests gives  more insight into the nature of the relationship. When both coefficients are significant, we expect linear relationships between the variables of both groups. However, it does not mean that there are only linear relationships and non-linear relationships between the variables may occur as well. 
When only the dCov coefficient is significant then we expect only non-linear relationships but no information is available about the nature of these relationships. 
One should also take into account that the RV and related coefficients  have had 30 years  of use and
the  development of  a large array of methods for dealing with multiway tables and heterogeneous multi-table data \citep{Kroonenberg08, Acar2009, Escou87, Lavit1994, dray07, FactoJSS, Pages14} that now allow 
the user to explore and visualize their complex multi-table data after assessing 
the significance of the associations.  Consequently, these coefficients have become part of a broader strategy 
for  analyzing heterogeneous data. 
We  illustrate in Section \ref{realdata} the importance of supplementing the coefficients and their test by graphical representations to investigate  the  significant relationships between blocks of variables.


\section{Beyond  Euclidean distances} \label{sec:beyond}

The RV coefficient and the dCov coefficient rely on  Euclidean distances (whether squared or not). In this section we focus on coefficients based on other distances or dissimilarities.

\subsection{The Generalized RV \label{GRVli}}

\citet{Minas13} highlighted the fact that the data are not always attribute data (with observations described by variables) but  can often be just distances or dissimilarity matrices, such as
  data from graphs such as social networks. They noted that the RV coefficient is only defined for Euclidean distances whereas other distances can be better fitted depending on the nature of the data. They referred for instance to the ``identity by state" distance or the Sokal and Sneath's distance which are well suited for specific biological data such known as SNP data. To overcome this drawback of the RV coefficient, they defined the generalized RV (GRV) coefficient as follows:

\begin{eqnarray}
 \mbox{GRV} (\bfX,\bfY)= \frac{<\bfC\boldsymbol{\Delta}^2_{\bfX}\bfC,\bfC\boldsymbol{\Delta}^2_{\bfY}\bfC>}{ \parallel \bfC\boldsymbol{\Delta}^2_{\bfX}\bfC\parallel
\parallel \bfC\boldsymbol{\Delta}^2_{\bfY}\bfC \parallel}
\label{GRV}
\end{eqnarray}
where  ${\Delta}_{\bfX}$ and ${\Delta}_{\bfY}$ being arbitrary dissimilarity matrices. 
The properties of their coefficient depend on the properties of the matrices $\bfC\boldsymbol{\Delta}^2_{\bfX}\bfC$ and $\bfC\boldsymbol{\Delta}^2_{\bfY}\bfC$. If both are positive semi-definite, then GRV varies between 0 and 1; if both have positive or negative eigenvalues then the GRV can take negative values but the value 1 can still be reached; if one is semi-definite positive and the other one not, the value 1 cannot be reached. 

To assess the significance of the GRV coefficient, they derived the first three moments of the coefficient based on \cite{Kaz}'s results and used the Pearson type III approximation of the permutation distribution. To deal with real data, they suggested  computing the GRV coefficient and using a test for different choices of distances for each matrix $\bfX$ and $\bfY$. Flexibility   is a strength 
here, since accommodating  different distances allows the user to see
different aspects of the data,
although this may cause disparities in power, the authors did suggest strategies for aggregating  results.




Note that  the dCov coefficient, although defined with Euclidian distances, could be extended in the same way to handle dissimilarity matrices.

\subsection{Kernel measures \label{noyau}}

The machine learning community has adopted similarity measures between kernels.
  Kernels are similarity matrices  computed from attribute data 
  or from non matrices data such as graphs, trees or rankings.  
A popular  similarity is the maximum mean discrepancy (MMD) between the joint distribution of two random variables and the product of their marginal distributions. This criterion introduced by \cite{Gretton05} is called the Hilbert Schmidt Independent Criterion (HSIC) and can be written as: 
\begin{eqnarray}
\mbox{HSIC}=\mbox{tr}(\bfK_{\bfX}\bfK_{\bfY})  
\label{HSIC}
\end{eqnarray}
with $\bf{K}_{\bfX}$ being a $n \times n$ kernel matrix for the first set (resp. $\bfK_{\bfY}$ for the second set). Note that this measure is an extension of the numerator of the RV coefficient \eqref{num_RV} since the RV numerator is the inner product between simple cross-product (kernel) matrices. 
\cite{Purdom06}  made the connection between the RV coefficient and the kernel literature by defining a  RV coefficient for the kernels. This is the correlation version of the HSIC \eqref{HSIC} which represents the covariance. \cite{Purdom06}  also  defined Kernel PCA and Kernel Canonical Correlation Analysis as maximizing the ``RV for kernels" between different kernels under constraints. 

Although the machine learning literature 
doesn't make connections with the RV literature, the supporting material
is very similar.
Tests of significance and asymptotic distributions under the null  are derived as similar to those covered in  Sections \ref{sec:asympt:rv} and \ref{sec:asympt:dcov}: $n \mbox{HSIC} ~ \sim \sum_{i=1}^{\infty} \sum_{j=1}^{\infty} \lambda_i \gamma_{j} Z_{ij}^2$
where $\lambda_i$ and $\eta_{j}$ are the eigenvalues of the operators. The empirical version of HSIC is also biased. \cite{SongL12} showed that the bias comes from the diagonal terms of the kernels and defined an unbiased estimator by removing these terms. 
\cite{Sejdinovic13} linked distance covariance coefficients and  HSIC  and  showed the equivalence between the  HSIC coefficient with specific choices of kernels and the dCov coefficient with specific power (Section \ref{sec:dcov_alpha}). 


Others related coefficients   are the kernel target alignment coefficient  \citep{Crist01}, many of
these coefficients are implemented in  MATLAB \citep{MATLAB}.





\subsection{Graph based measures \label{graph}}

Early versions of association measures were related to closeness between graphs \citep{Barton1962}. 
More recently, \citet{Frie83}  defined a very useful such coefficient. 
Their method supposes
sets of interest (either the two matrices $\bfX$ and $\bfY$ with attribute data or two matrices of dissimilarities) represented by two complete graphs where each observation is a node (there are $n$ nodes) and the $(n(n-1)/2)$ edges are weighted by a dissimilarity (the Euclidian distance can be used as well). Then, they built two spanning subgraphs, usually the $k$ nearest-neighbor (KNN) graph where  an edge 
is built between a node and its $k$ neighbors (the other alternative is the $k$ minimal spanning tree). The test statistic is the number of edges common to the two graphs. When many observations  connected in
one graph are also connected in the other, this measure of association is high. The main feature of such a measure is that the larger distances are not considered which ensures the test to be powerful against non-monotone alternatives. However, we may expect less power to detect monotone relationships
than the coefficients studied in Section \ref{sec_RV} and \ref{sec:dcov}. 
\citet{Frie83} also derived the first two moments of the permutation distribution under the null hypothesis of independence and detailed the situations where an asymptotic normal approximation can be considered.
The power of the tests depend  on the choice of dissimilarities (even if it robust enough since it depends only on the rank order of the edges) as well as on the number $k$ for the KNN approach. They also highlighted that ``\textit{values significant should be used to signal the need to examine the nature of the uncovered relationship, not as a final answer to some sharply defined question.}''
This coefficient was one of the first that allowed  detection of non-linear relationships. 
We will see in Section \ref{sec:microarray} that the k minimum spanning version is less powerful than the k-nearest
neighbor based coefficient.

\citet{Heller20123097} defined a related approach (without actually referring to \citet{Frie83}'s paper). Their test is also based on the use of minimal spanning tree but the rationale is to state that under the null, close observations in one graph are no longer close in the other graph and thus their ranks are randomly distributed.  Using similar simulations as those in Section \ref{simu}, they showed that their approach has better power than the one based on dCov. 

\subsection{The Mantel coefficient \label{sec:mantel}}

The Mantel \citep{Mant67, Leg10} coefficient, one of the earliest version of association measures, is probably also the most popular now, especially in ecology \citep{SokalSneath}. 
Given arbitrary dissimilarity matrices, it is defined as: 

\begin{eqnarray}
\mbox{r}_{\rm m}(\bfX,\bfY)= \frac{\sum_{i=1}^n \sum_{j=1,j \neq i}^n (d_{ij}^{\bfX}-\bar d^{\bfX})(d_{ij}^{\bfY}-\bar d^{\bfY})}{\sqrt{\sum_{i,j,j \neq i}(d_{ij}^{\bfX}-\bar d^{\bfX})^2 \sum_{i,j,j \neq i}(d_{ij}^{\bfY}-\bar d^{\bfY})^2}}, \nonumber
\end{eqnarray}
with $\bar d^{\bfX}$ (resp $\bar d^{\bfY}$) the mean of the upper diagonal terms of the dissimilarity matrix associated to $\bfX$ (resp. to $\bfY$). This is the correlation coefficient between the vectors gathering the upper diagonal terms of the dissimilarity matrices.  The main difference between the Mantel coefficient and the others such as the RV or the dCov is the absence of double centering. Its significance is assessed via a permutation test. The coefficient and its test are implemented in several R packages such as {\tt ade4} \citep{dray07},  {\tt vegan} \citep{vegan} and  {\tt ecodist} \citep{ecodist07}. 

Due to its popularity, many studies suggesting new coefficients often compared their performance to Mantel's. \citet{Minas13} showed that the Mantel test is less powerful than the test based on the GRV coefficient \eqref{GRV} using simulations. In the same way, \citet{Omelka2013} underlined the superiority of the dCov test over the Mantel test.
However, despite its widespread use, some of the properties of the Mantel test are unclear and recently its utility questioned \citep{Omelka2013}.
\citet{Leg10} showed that the Mantel coefficient is not equal to 0 when the covariance between the two sets of variables is null and thus can't be used to detect linear relationships. Non-linear relationships can be detected, there are not yet clear theoretical results  available to determine when. 

Nevertheless, the extensive use in ecology and spatial statistics has led to a large number of extensions of the Mantel coefficient. \citet{Smouse}
proposed a generalization  that 
 can account for a third type of variable, \textit{i.e.}
allowing for partial correlations. Recently, the lack of power and high type I  error rate
for this test has been noted, calling into doubt the validity of its use \citep{Guillot2013}.
\citet{Szekely13c} also considered this extension to a partial correlation coefficient based on dCov.

\section{Real data analysis} \label{realdata} 
 
Since the dCov coefficient has the advantage over the other
 coefficients such as the RV or the Procrustes coefficients of measuring departure from independence, it would be worthwhile for the ecologists, food-scientists and other scientists in applied fields to try the dCov on their data. 
 
 In this section, we illustrate the use of the coefficients and their test on different real data sets coming from different fields. We emphasize the complementarity of the different coefficients as well as the 
 advantage of providing follow-up  graphical representations. Many multi-block methods  that use
 the earlier RV could be adapted to incorporate the dCov coefficient as well. 
 (Code to reproduce the examples is available as supplementary material). 

\subsection{Sensory analysis}

\subsubsection{Reproducibility of tasting experiments.}

Eight wines from Jura (France) have been evaluated  by twelve panelists in the following way.
Each panelist tasted the wines and positioned them on
a 60$\times$40 cm sheet of paper in such a way that two wines are close if they
seem similar to the taster, and farther apart if they seem different. Then, the product coordinates are collected
in a 8 $\times$ 2 matrix. This way of collecting sensory data is named ``napping" \citep{Pag2005} and encourages spontaneous description.  
The 8 wines were evaluated during 2 sessions (with an interval of a few days). Thus, there are as many
matrices as there are couple taster-sessions (24 = 12 $\times$  2).  As
with any data collection procedures, the issue of repeatability arises here. Is
the products configuration given by a taster roughly the same from one session
to the other? In other words, do they perceive the wines in a same way during
the two sessions? This question was addressed in \cite{Josse08} by using the RV  between the configurations obtained during sessions 1 and 2 for all the panelists, here we add
 the dCov coefficient with different powers on the distances. Results are gathered in Table 1.

\begin{table}[ht]
\centering
\begin{tabular}{cccccccc}
  \hline
 & RV & RV$_p$ & dCor  & dCov$_p$ & dCov$_p$0.1 & dCov$_p$0.5 & dCov$_p$1.5 \\ 
  \hline
1 & 0.55 & 0.04 & 0.10 & 0.09 & 0.16 & 0.13 & 0.13 \\ 
  2 & 0.22 & 0.60 &  0.72 & 0.76 & 0.84 & 0.81 & 0.81 \\ 
  3 & 0.36 & 0.16 &  0.68& 0.32 & 0.55 & 0.43 & 0.44 \\ 
  4 & 0.13 & 0.68 &  0.84 & 0.76 & 0.51 & 0.65 & 0.65 \\ 
  5 & 0.64 & 0.02 &  0.01 & 0.02 & 0.04 & 0.03 & 0.03 \\ 
  6 & 0.14 & 0.56 &  0.54 & 0.75 & 0.83 & 0.81 & 0.81 \\ 
  7 & 0.79 & 0.01 &   0.91& 0.01 & 0.01 & 0.01 & 0.01 \\ 
  8 & 0.06 & 0.82 &  0.81 & 0.76 & 0.65 & 0.70 & 0.70 \\ 
  9 & 0.49 & 0.04 &  0.28 & 0.11 & 0.28 & 0.25 & 0.25 \\ 
  10 & 0.28 & 0.29  & 0.29 & 0.24 & 0.17 & 0.20 & 0.20 \\ 
  11 & 0.22 & 0.40  & 0.39 & 0.26 & 0.19 & 0.23 & 0.22 \\ 
  12 & 0.19 & 0.54  & 0.58 & 0.55 & 0.58 & 0.57 & 0.56 \\ 
   \hline
\end{tabular}
\label{table:napping}
\caption{Coefficients of association and tests between the configuration of the 12 tasters obtained during session 1 and  session 2: RV coefficient and its $p$-value  RV$_p$, dCor coefficient and its $p$-value and the $p$-values associated with the dCov test with power $\alpha$ equals to  0.1, 0.5 and 1.5 on the distances. RV test is performed using Pearson's approximation and the other tests with 1000 permutations.} 
\end{table}

\begin{figure}[!ht]
\includegraphics[width=0.8\textwidth,height=2.3in]{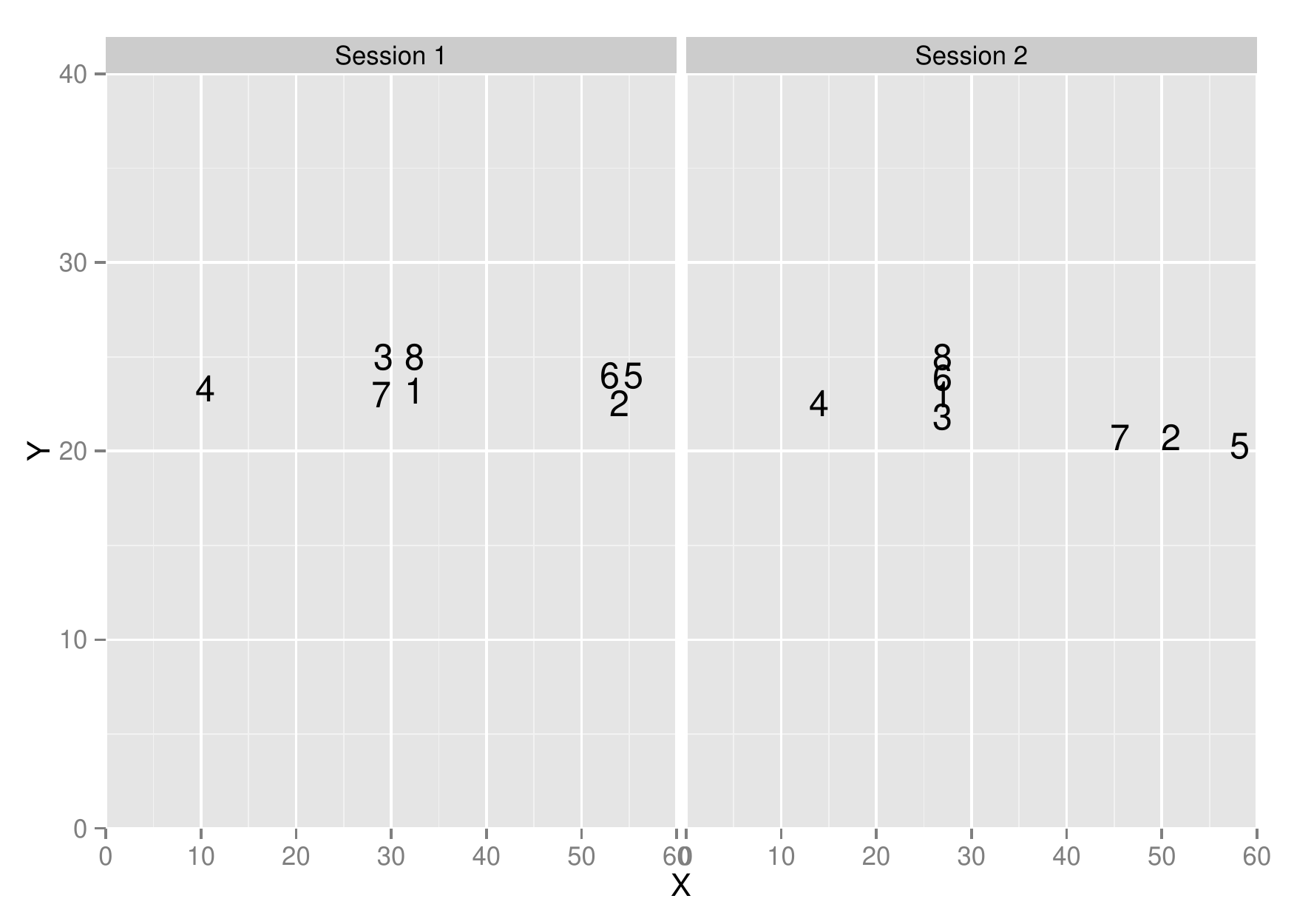}
\caption{Representation of the 8 wines on the $40 \times 60$ sheet of paper given by the panelist 9 during session 1 and 2. }
\label{fig:pane9}
\end{figure}

\begin{figure}[!ht]
\includegraphics[width=0.8\textwidth,height=2.3in]{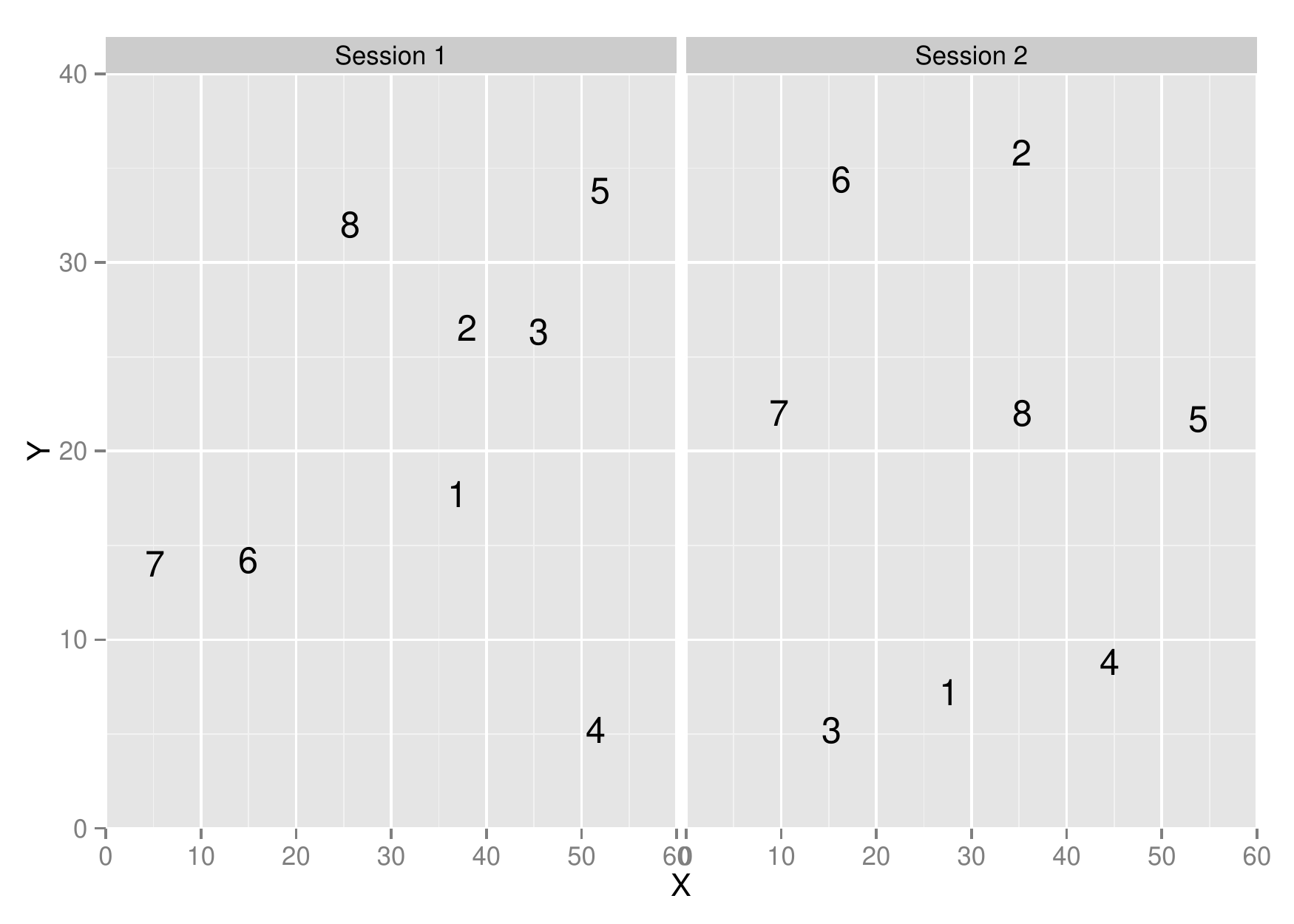}
\caption{Representation of the 8 wines on the $40 \times 60$ sheet of paper given by the panelist 1 during session 1 and 2. } 
\label{fig:pane1}
\end{figure}

\begin{figure}[!ht]
\centering
\includegraphics[width=0.6\textwidth,height=2.3in]{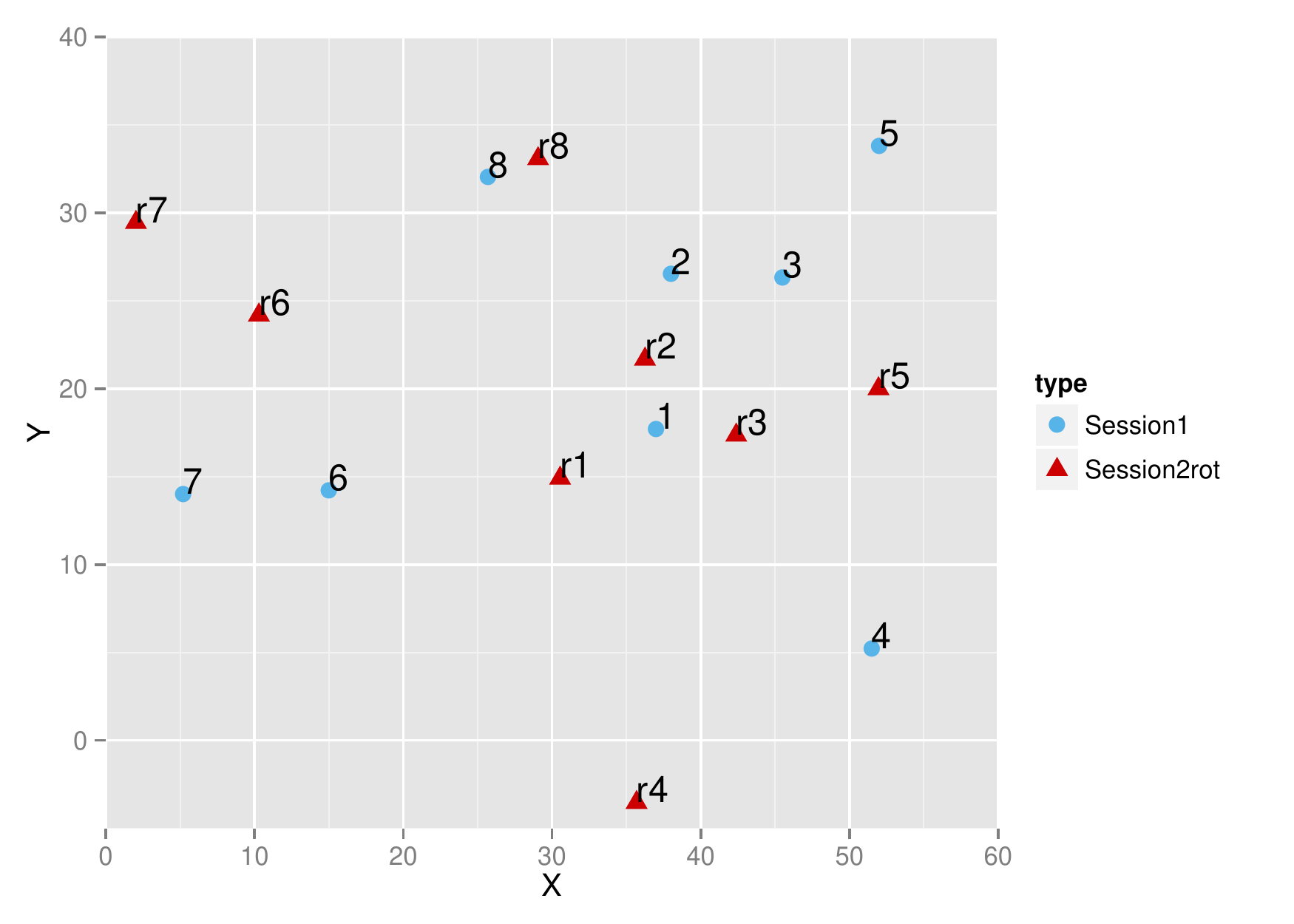}
\caption{Representation of the rotated configuration of the session 2 (red triangles) onto the session 1's configuration for panelist 1.} 
\label{fig:pane1rot}
\end{figure}

\begin{figure}[!ht]
\includegraphics[width=0.8\textwidth,height=2.3in ]{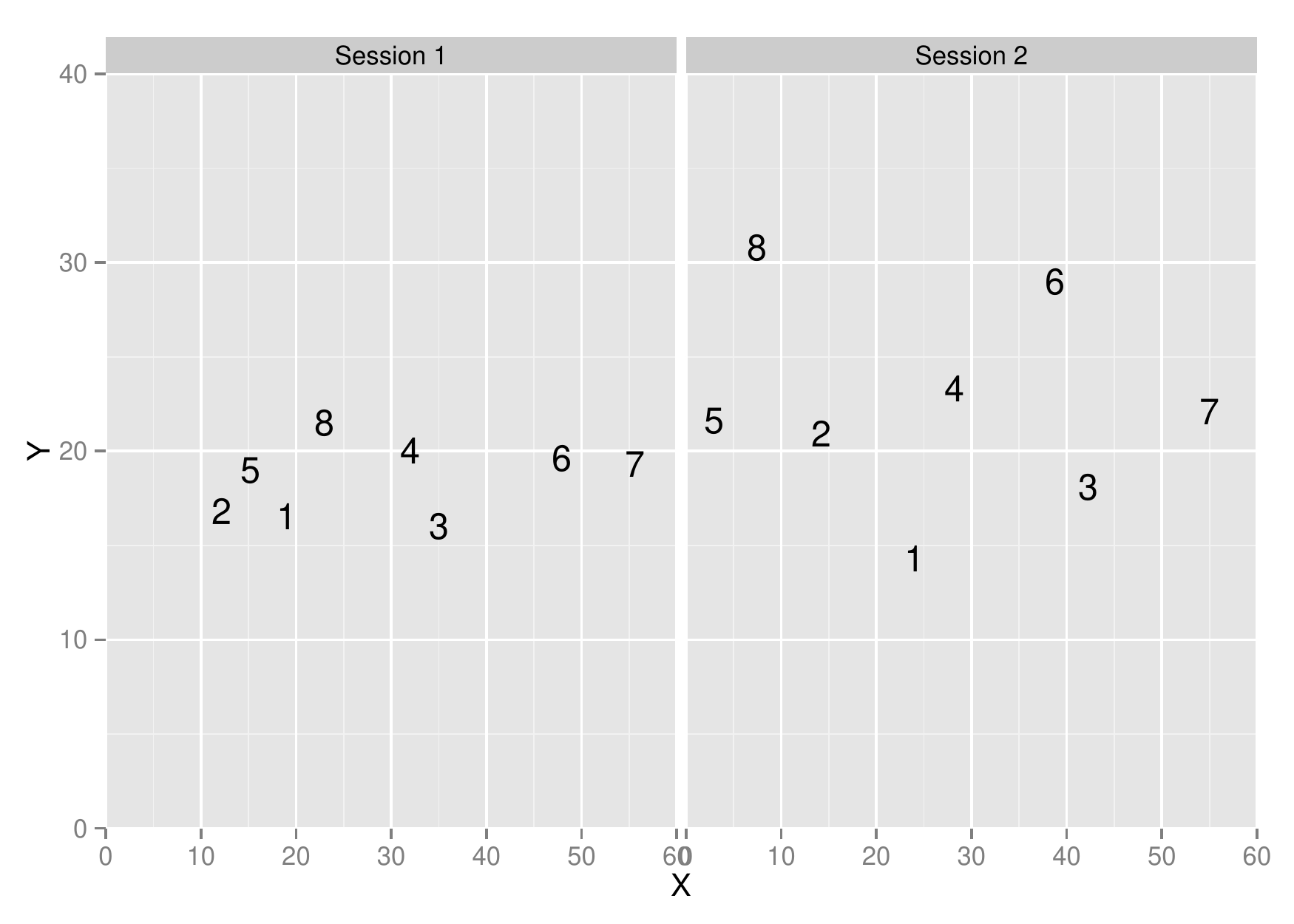}
\caption{Representation of the 8 wines on the $40 \times 60$ sheet of paper given by the panelist 7 during session 1 and 2.}
\label{fig:pane7}
\end{figure}

The methods show that tasters 5 and 7 are repeatable. 
For tasters 1 and 9, only the RV coefficient rejects the null. Figures \ref{fig:pane9} and \ref{fig:pane1} give their representation during the first and second sessions. Taster 9 distinguished 3 clusters of wines but switched the wines 6 and 7  from one session to the other. It is more difficult to understand why the RV coefficient is significant when inspecting the configurations given by taster 1. However, since the RV is invariant by rotation, we rotated the second configuration onto the first one on Figure \ref{fig:pane1rot}.  The pattern looks more similar with wines 6 and 7 quite close and the wine 4 far from the others. 
Figure \ref{fig:pane7} gives the representation provided by taster 7  to show a case with a consensus between the tests. 
On this real data set, it is impossible to know the ground truth but the RV test spotlights two panelists that may be considered as reliable. 

\subsubsection{Panel comparison. \label{sec:panel}}

Six French chocolates were evaluated by 7 panels of 29 judges
who grade 14 sensory descriptors such as bitterness, crunchy, taste of caramel, etc. For each panel, the data matrix is  of size $6 \times 14$ and each cell corresponds to the average of the scores given for one chocolate on a descriptor by the judges  (ranging from 1 for not bitter to 10 for very bitter for instance). One aim of the study was to see if the panels produce 
concordant descriptions of the products. Tables 2 and  3 
show the matrices of  RV and dCor coefficients. 
All the coefficients are very high and are highly significant. 
\begin{table}[ht]
\begin{center}
\begin{tabular}{rrrrrrrr}
  \hline
 & 1 & 2 & 3 & 4 & 5 & 6 & 7 \\ 
  \hline
1 & 1.000 & 0.989 & 0.990 & 0.984 & 0.985 & 0.995 & 0.993 \\ 
  2 &  & 1.000 & 0.992 & 0.991 & 0.993 & 0.996 & 0.997 \\ 
  3 &  &  & 1.000 & 0.995 & 0.992 & 0.996 & 0.997 \\ 
  4 &  &  &  & 1.000 & 0.983 & 0.993 & 0.993 \\ 
  5 &  &  &  &  & 1.000 & 0.994 & 0.997 \\ 
  6 &  &  &  &  &  & 1.000 & 0.999 \\ 
  7 &  &  &  &  &  &  & 1.000 \\ 
   \hline
\end{tabular}
\label{table:rvcoeff_choco}
\end{center}
\caption{RV coefficients between the matrices products-descriptors provided by the 7 panels.} 
\end{table}

\begin{figure}[!ht]
\includegraphics[width=\textwidth]{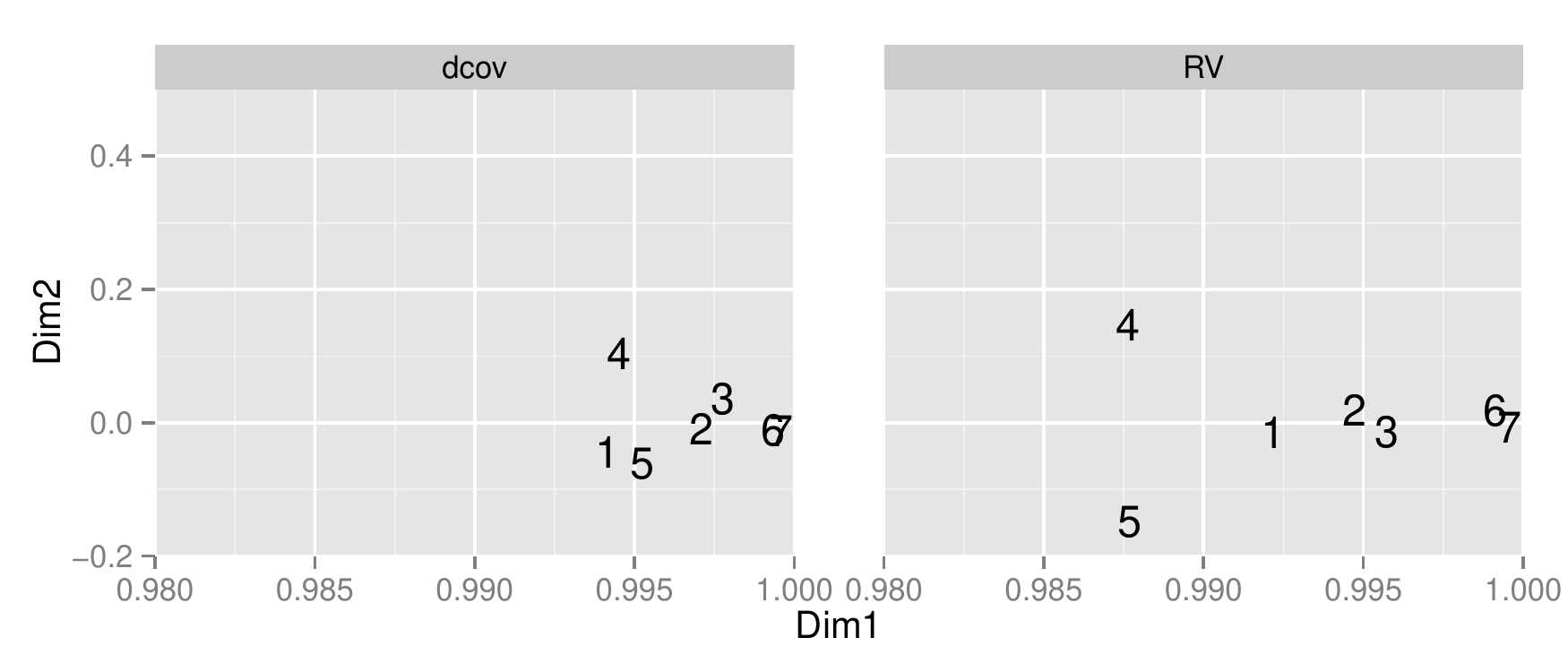}
\caption{Graphical representation of the proximity between panels with the proximity defined  with the dCor coefficient (on the left) and with the RV coefficient (on the right).}
\label{fig:inter_choco}
\end{figure}

\begin{table}[ht]
\begin{center}

\begin{tabular}{rrrrrrrr}
  \hline
 & 1 & 2 &  3 &  4 & 5 &  6 &  7 \\ 
  \hline
1 & 1.000 & 0.986 & 0.983 & 0.974 & 0.977 & 0.991 & 0.991 \\ 
  2 &  & 1.000 & 0.984 & 0.981 & 0.978 & 0.996 & 0.995 \\ 
  3 &  &  & 1.000 & 0.984 & 0.987 & 0.993 & 0.994 \\ 
   4 &  &  &  & 1.000 & 0.956 & 0.988 & 0.986 \\ 
   5 &  &  &  &  & 1.000 & 0.983 & 0.989 \\ 
   6 &  &  &  &  &  & 1.000 & 0.999 \\ 
   7 &  &  &  &  &  &  & 1.000 \\ 
   \hline
\end{tabular}
\label{table:dcovcoeff_chocoAA}
\end{center}
\caption{dCor coefficients between the matrices products-descriptors provided by the 7 panels.} 
\end{table}
After performing this step, the analysis of the RV matrix is undertaken by doing a multi-block method such as STATIS \citep{Escou87}. The  rationale of STATIS is to consider the matrix of RV's as a matrix of inner products. Consequently,  an Euclidean representation of the inner products  can be made in a lower-dimensional space by performing the eigenvalue decomposition of the matrix. This is the first step of STATIS named the  ``between-structure" which  produces a graphical representation of the proximity between tables. 
This can be quite useful when there are many blocks of variables. This is equivalent to performing multidimensional scaling (MDS or PCoA)  \citep{Gower66}  on the associated distance matrix. The same reasoning is valid for a
matrix of dCor coefficients and thus we use this  same approach on the dCor matrix.  
Figure \ref{fig:inter_choco} is the result of such an analysis and shows that there is strong consensus between the  description of the chocolates provided by the 7 panels since the 7 panels are very close.

\begin{figure}[!ht]
\includegraphics[scale=0.3]{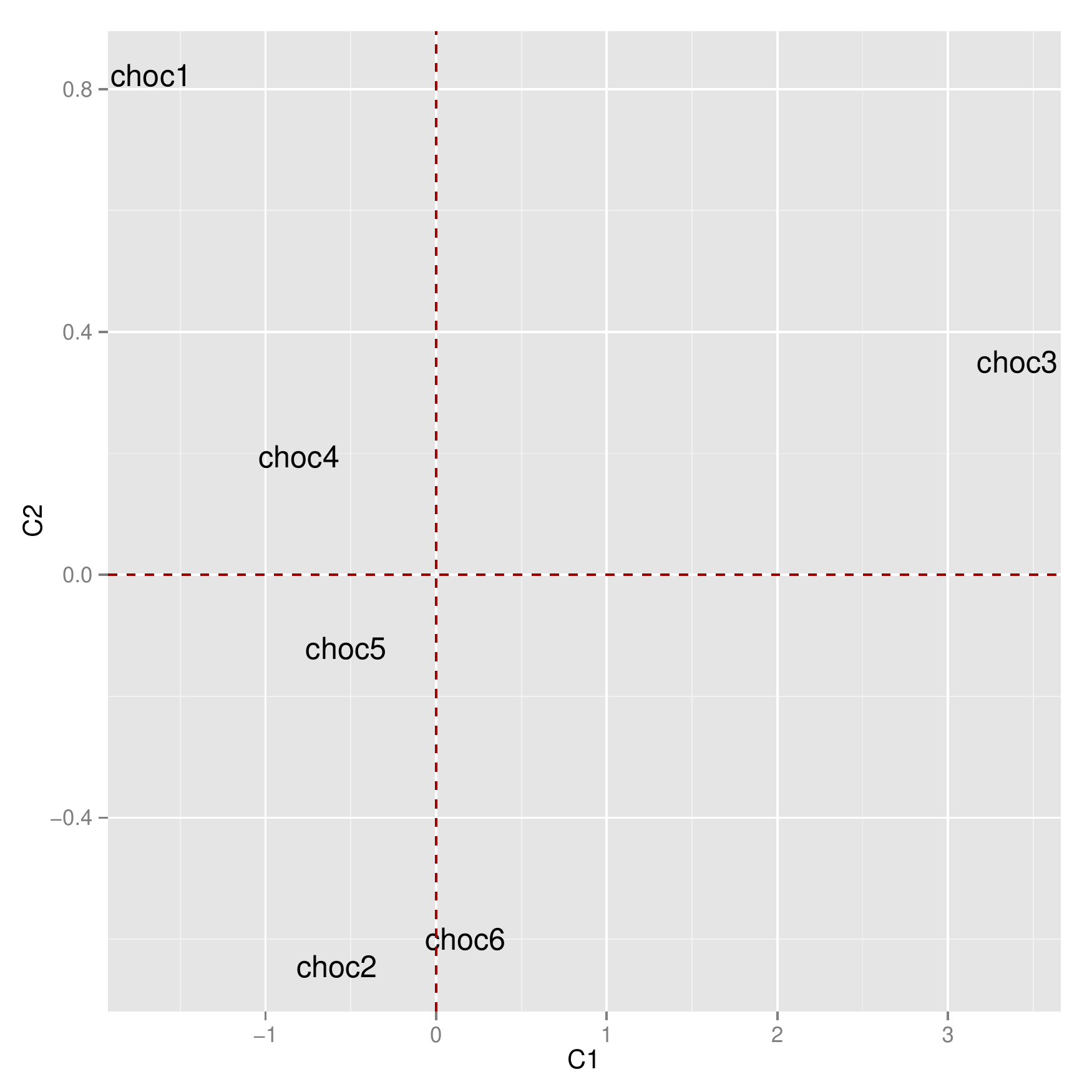}
\caption{Representation of the STATIS compromise.}
\label{fig:statis_representation}
\end{figure}

The STATIS method goes deeper by inspecting what is common between the 7 panels (called the  ``compromise" step) and then by looking at what is specific to each panel (called the  ``within-structure" step). 
This could also be useful for the dCor coefficient.
The ``compromise" representation is obtained by looking for a similarity matrix $\bar \bfW$ which is the more related to all the inner product matrices (here $K$=7) in the following sense: \\$\bar \bfW=\mbox{argmax}_{\bfW^{\star}=\sum_{k=1}^{K} \gamma_k \bfW_k; \sum_{k} \gamma_{k}^2=1}\sum_{k=1}^{K}< \bfW^{\star},\bfW_k>^2$.
The weights $\gamma_k$ are given by the first eigenvector of the RV matrix and are positive since all the elements of the RV matrix are positive (using the Frobenius theorem). Then an Euclidean representation of the compromise object $\bar \bfW$ is also obtained by performing the eigen decomposition and is given Figure \ref{fig:statis_representation}. It shows that all the 7 panels distinguished chocolate 3 from the others. We do not detail the sequel of the analysis which would consist in looking at why the chocolate 3 is so different from the other, etc.

 We could also note that one could also consider the analogous of STATIS for kernels and get as a compromise kernel a linear combination of kernels with optimal weights.

\subsection{Microarray data \label{sec:microarray}}

We continue the example discussed in the introduction on the 43 brain tumors described with expression data (356 variables) and CGH data (76 variables). 
\subsubsection{Distance based coefficients}
We compare the two different types of information
first by computing the RV coefficient. A high value would indicate that when tumors are similar from the point of view of the transcriptome they are also similar from the point of view of the genome.  
The RV coefficient is equal to 0.34. Section \ref{sec:rv:biais} showed the importance of computing a debiased version of the coefficient especially when dealing with large data. The debiaised RV described Section \ref{sec:rv:biais} is not implemented but we debiased it by removing its expectation under the null defined equation \eqref{expectation} which is equal to ${\mathbb E}_{H_0}(\mbox{RV})=0.16$. 
The dCor coefficient is equal to 0.74 and its debiaised version dCor$^{*}$ to 0.28.
These coefficients are significant. However a significant coefficient may cover very different situations. 

To put into perspective these values, let us consider a simulation with two data sets of the same size $\bfX_{43 \times 68}$ and $\bfY_{43 \times 356}$ where all the variables are generated under a normal distribution with expectation equal to zero and variance equal to one.  
The first row of Table 4 corresponds to the median over 1000 simulations of the results obtained under this null setting. 	Note that both the RV and the dCor coefficients are very high but their debiased version are roughly equal to zero hence the importance of looking at the debiased coefficients.  Such large values can be explained because of the size of the data and because all the variables within each table are uncorrelated.
95\% of the tests are not significant which was also expected (here we report only the median of the $p$-value which is around 0.5 for all the tests).  
\begin{table}[ht]
\label{table:simu_gene}
\begin{center}
\begin{tabular}{c c c c c c c c c c c}
  \hline
 & RV & 
RV$^{*}$& 
RV$_p$ & 
dCor &
 dCor$_p$  &
 dCor$^{*}$ &
 dCor$^{*}_p$ & 
 dCov$_p$0.1 &
 dCov$_p$0.5 &
 dCov$_p$1.5  \\ 
  \hline
1 & 0.740 & 0.002 & 0.465  & 0.957 & 0.495  & 0.003 & 0.468 & 0.514 & 0.541 & 0.507 \\ 
 2 & 0.715& 0.014 & 0.046  & 0.950 & 0.048& 0.055 & 0.053 & 0.057 & 0.056 & 0.045 \\ 
 3&0.728  &  0.001&    0.452  &     0.954 &    0.451& 0.004&    0.448 &   0.437&    0.466 &   0.449 \\  
   \hline
\end{tabular}
\end{center}
\caption{Relationship between simulated data $\bfX_{43 \times 68}$ and $\bfY_{43 \times 356}$ using the different coefficients and tests: RV coefficient,  debiased RV$^{*}$ and its $p$-value RV$_p$ (using Pearson's approximation); dCor coefficient and its $p$-value dCor$_p$ (using permutation tests); unbiased dCor$^{*}$ coefficient and its $p$-value dCor$_p^{*}$ (using the asymptotic test);  dCov tests with power 1, 0.1, 0.5 and 1.5 on the distances. The dCov tests are performed with 1000 permutations. Row 1: $\bfX$ and $\bfY$ are independent. Row 2:  linear relationships $\bfX_j=2 \bfY_j+ \varepsilon$, for $j=1,2,3$. Row 3: non-linear relationships $\bfX_j=\mbox{log}(\bfY_j^2)+ \varepsilon$, for $j=1,...,68$.} 
\end{table}
Row 2 of Table 4 corresponds to a case where 3 variables of $\bfX$ are generated  as  $\bfX_j=2 \bfY_j+ \varepsilon$, for $j=1,2,3$ and $\varepsilon$ a gaussian noise with a small variance (0.02). Note that the tests are significants despite the fact that only few variables are related. However, the debiased coefficients are small. 
Finally, row 3 of Table 4  corresponds to a case where all the variables of $\bfX$ are generated  as  $\bfX_j=\mbox{log}(\bfY_j^2)+ \varepsilon$, for $j=1,...,68$. It shows that  non-linear relationships are harder to discover since the medians of the 1000 $p$-values of the dCor tests with permutations and with the asymptotic test are around 0.45. With these simulations (keeping $p=68$ and $q=356$), significant results are obtained when increasing the sample size $n$ to 1000.

\subsubsection{Graph based coefficients}
\begin{figure}[!hbt]
\centerline{\includegraphics[width=0.80\textwidth,height=0.3\textwidth]{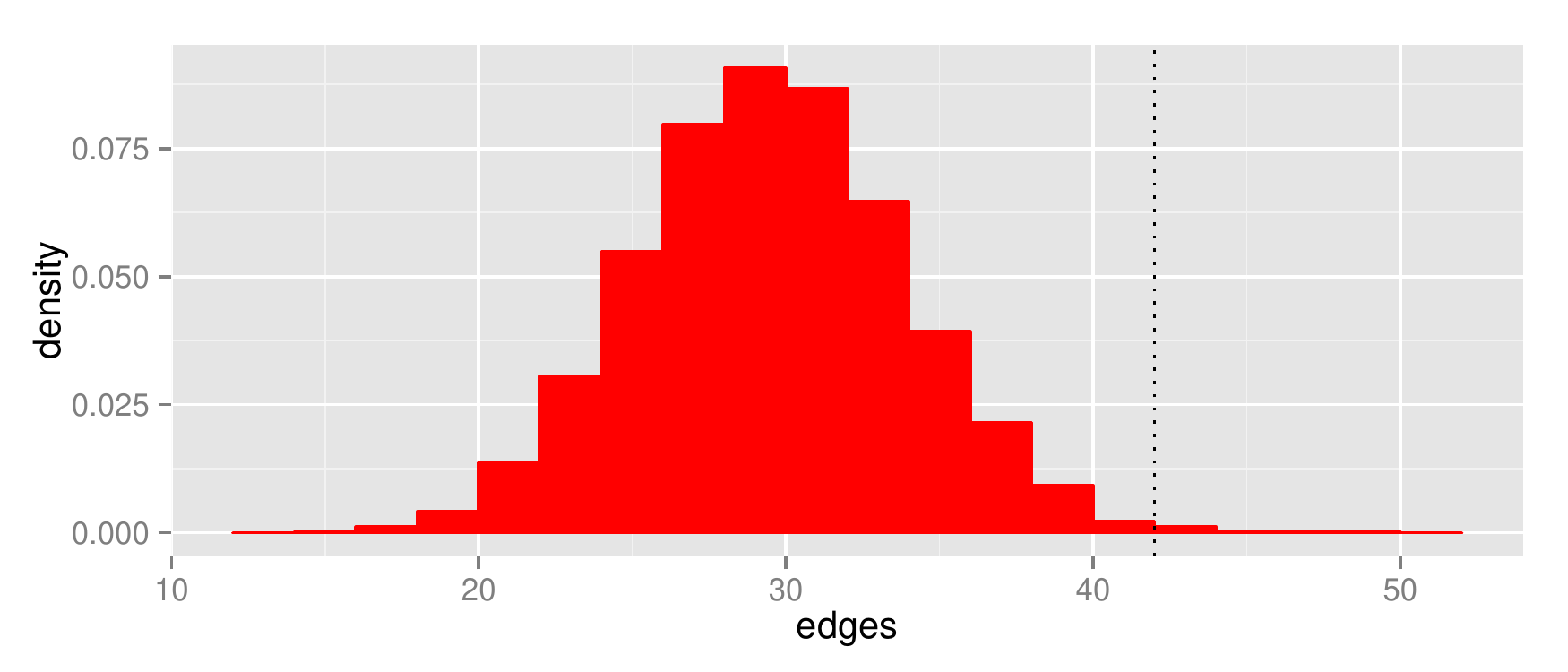} }
\caption{Histogram of the permutation distribution of Friedman and Rafsky's k nearest-neighbor  graphs' common
edges with k=5, the observed value was 42 for the original data.}
\label{fig:knnhist}
\end{figure}

Here we  computed 
coefficients following \citet{Frie83} (described Section \ref{graph})
using both the minimum spanning trees and the
k nearest-neighbor trees. The former showed very little association and seems to
have very little power in high dimensions, the two minimum spanning trees only had
three edges in common out of 42.
However, as shown in Figure \ref{fig:knnhist}, the k nearest-neighbor version (with k=5)
is significant with a $p$-value smaller than 0.004.

\subsubsection{Graphical exploration of associations}
\begin{figure}[!hbt]
\centerline{\includegraphics[width=0.50\textwidth]{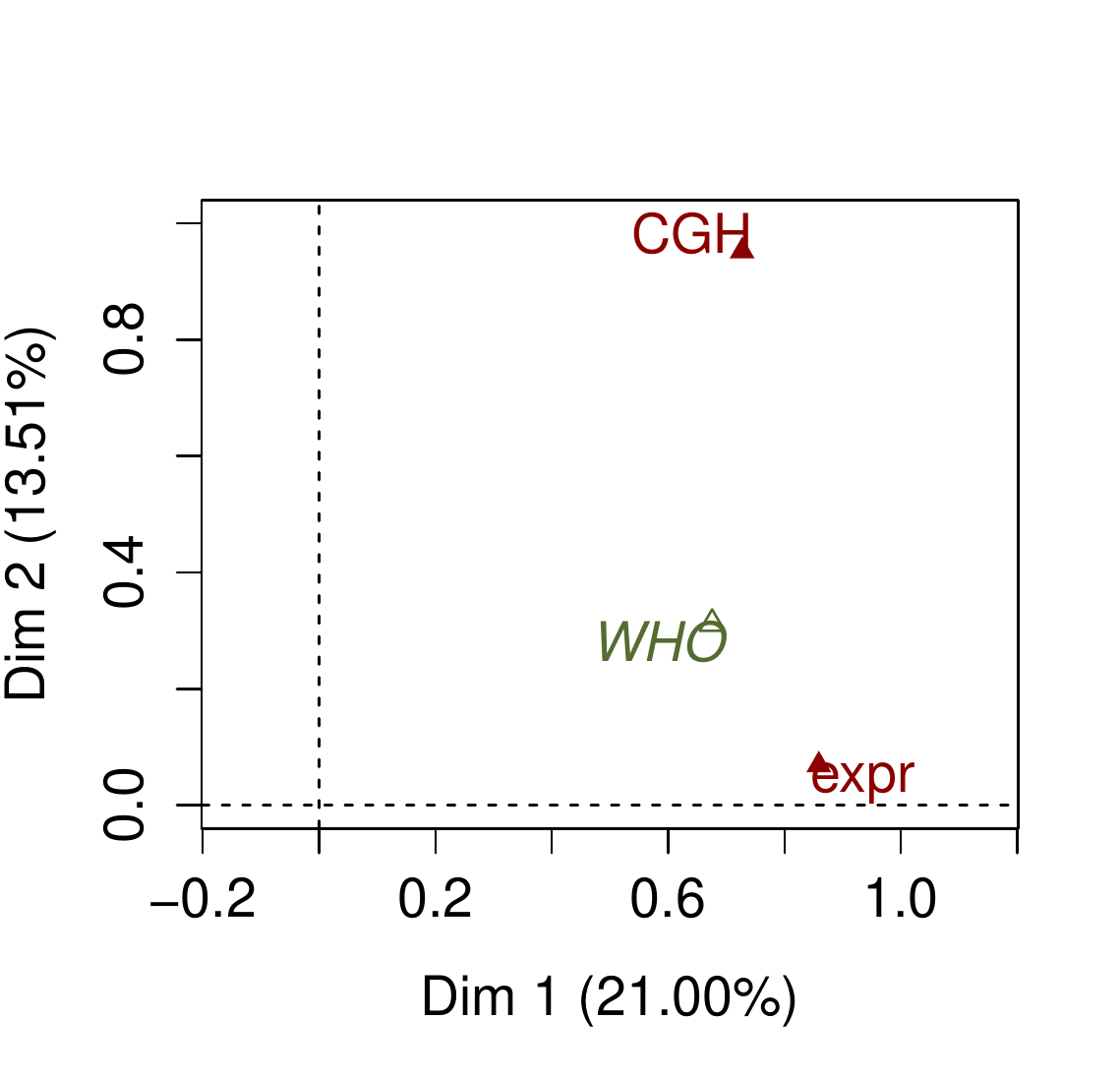}  ~ \includegraphics[width=0.50\textwidth]{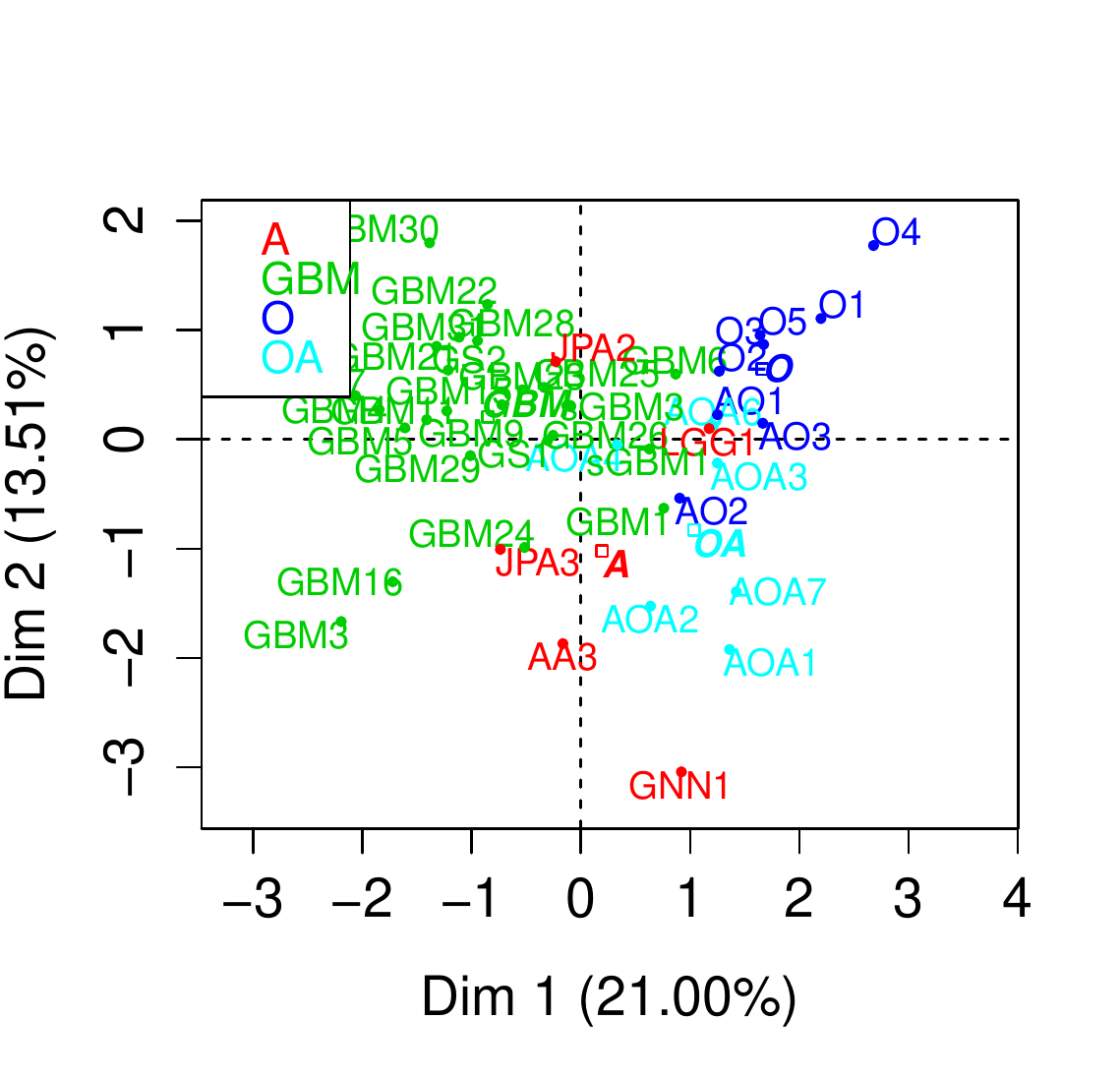} }
\caption{MFA groups representation (left) and  compromise representation of the tumors (right).}
\label{fig:group_genes}
\end{figure}

\begin{figure}[!hbt]
\centerline{\includegraphics[width=0.45\textwidth]{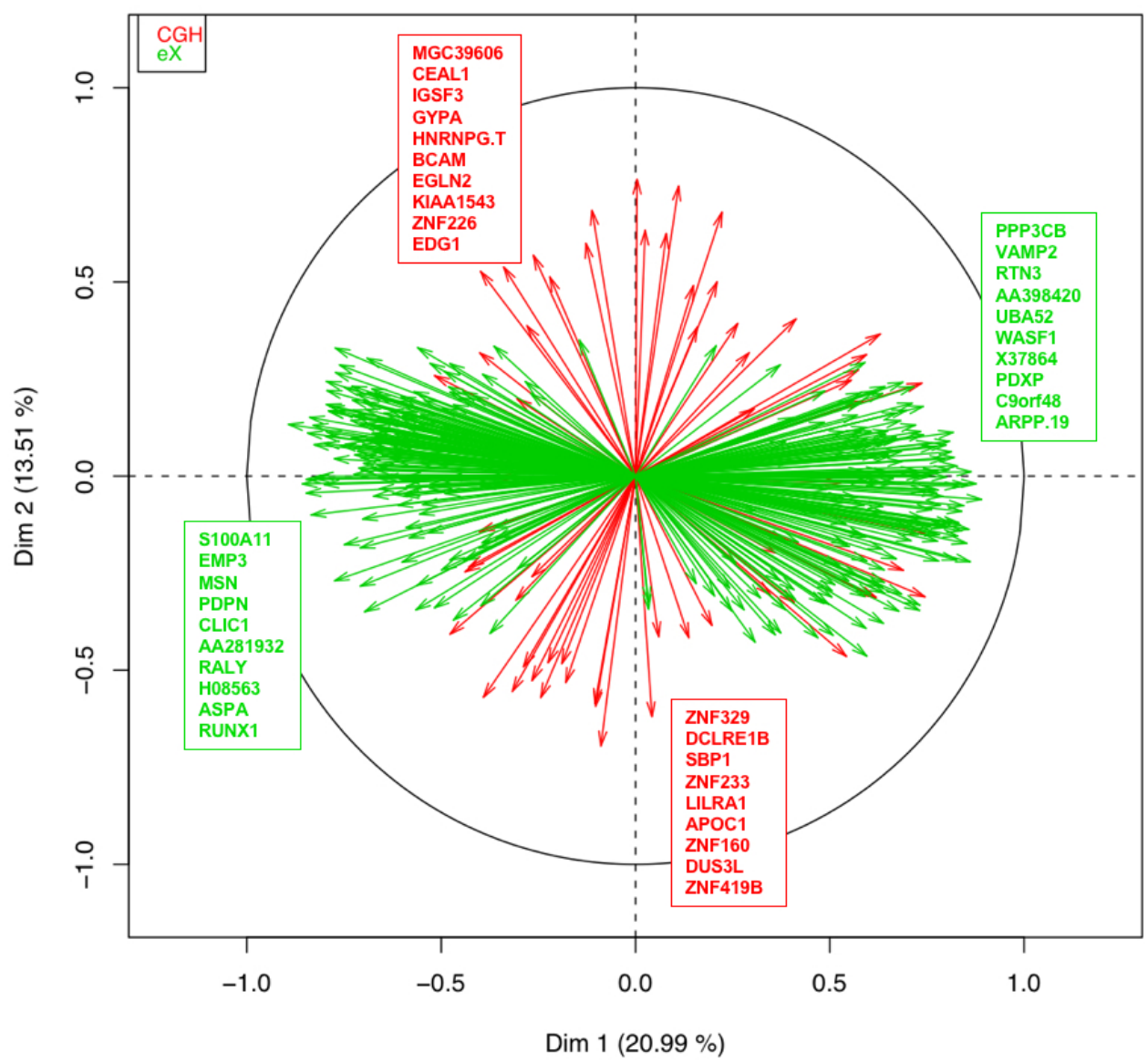} ~ \includegraphics[width=0.45\textwidth]{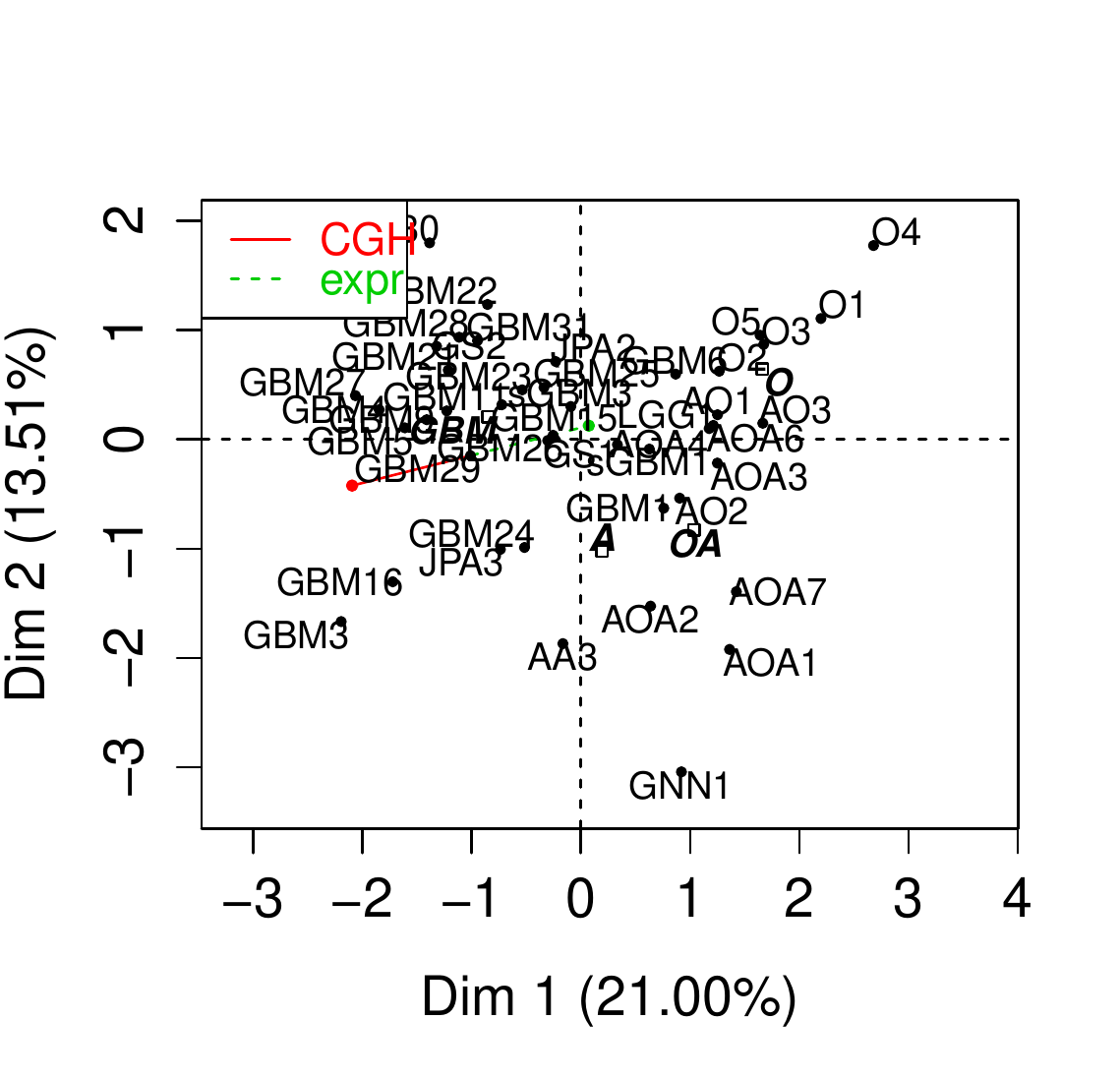}}
\caption{IMFA variables representation (left) and a ``partial" sample (right).}
\label{fig:ind_var_genes}
\end{figure}

The previous results and simulations
point to the existence of some linear relationships between the variables of both groups. 
To describe and visualize the associations, different multi-block methods such as STATIS are available in the literature \citep{Kroonenberg08} and we focus here on multiple factor analysis (MFA) described in \cite{Pages14}. This method uses the $L_g$ coefficient described Section \ref{sec:lgcoeff}. The $L_g$ coefficient for the expression data is equal to 1.09 whereas it is 2.50 for the CGH data which means that the expression data may be uni-dimensional whereas the CGH data is more multi-dimensional. MFA gives as an output Figure~\ref{fig:group_genes} on the left which is the equivalent of the ``between-structure"  step of Section  \ref{sec:panel}. 
Here, the coordinates of the groups corresponds to the values of the $L_g$ between the dimensions of the ``compromise" and each group. Thus, it shows that that the  first dimension is common to both groups whereas the second dimension is mainly due to the group CGH. We are also able to say that this first dimension is close to the first principal component of each group since the values of the $L_g$ are close to one (as explained in Section \ref{sec:lgcoeff}). 
Figure~\ref{fig:group_genes} on the right  is the equivalent of the ``compromise"  step of Section \ref{sec:panel} and shows that the first dimension of variability opposes the glioblastomas tumors to the lower grade tumors and that the second dimension opposes tumors O to the tumors OA and A. 
Since as mentioned previously, the first dimension is common to both groups of variables, it means that both the expression data and the CGH data permits to separate the glioblastomas to the other tumors. On the other hand, only the CGH data permits to see differences between the tumor O and the tumors OA and A.  Thus, it shows what is common and what is specific to each group. Figure~\ref{fig:ind_var_genes} on the left is the correlation circle to study the correlation between all the variabes and shows that the expression data is much more one-dimensional whereas the CGH data is represented at least on two dimensions (red arrows are hidden by the green arrows) which was expected due to the values of the $L_g$ coefficients. 
This method also allows  to compare the information of both groups at the observation level with the  ``partial" representation
represented  Figure~\ref{fig:ind_var_genes} on the right. The tumor GBM29 is represented using only its expression data (in green) and using only its CGH data (in red). The black dot is at the barycenter of both red and green points and represents the tumor GBM29 taking into account all the data. 
  This tumor is particular in the sense that when taking its CGH data, this individual is on the side of the dangerous tumors (small coordinates on the first axis) whereas it is on the side of the other tumors when considering its expression data (positive coordinates on the first axis). There is no consensus for this individual between the two sources of information and it may require more investigation to understand why.
More details about the method and rules of interpretation can be found in  \cite{Pages14}. Note that we only inspect the linear relationships and potential non-linear relationships highlighted by the dCov coefficient are not studied. 

\subsection{Morphology data set}

\begin{figure}[!hbt]
\centerline{\includegraphics[width=0.47\textwidth]{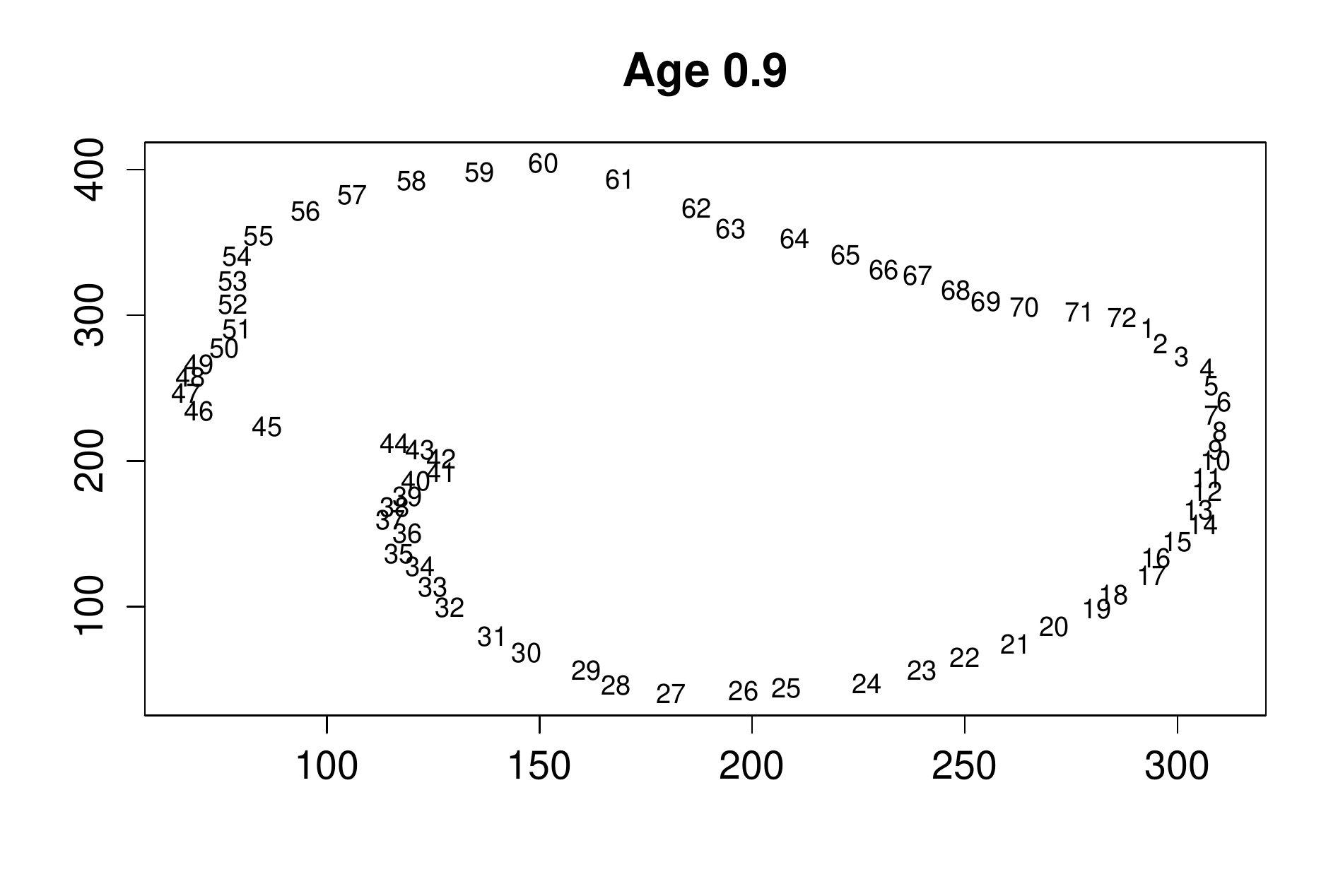} ~ \includegraphics[width=0.47\textwidth]{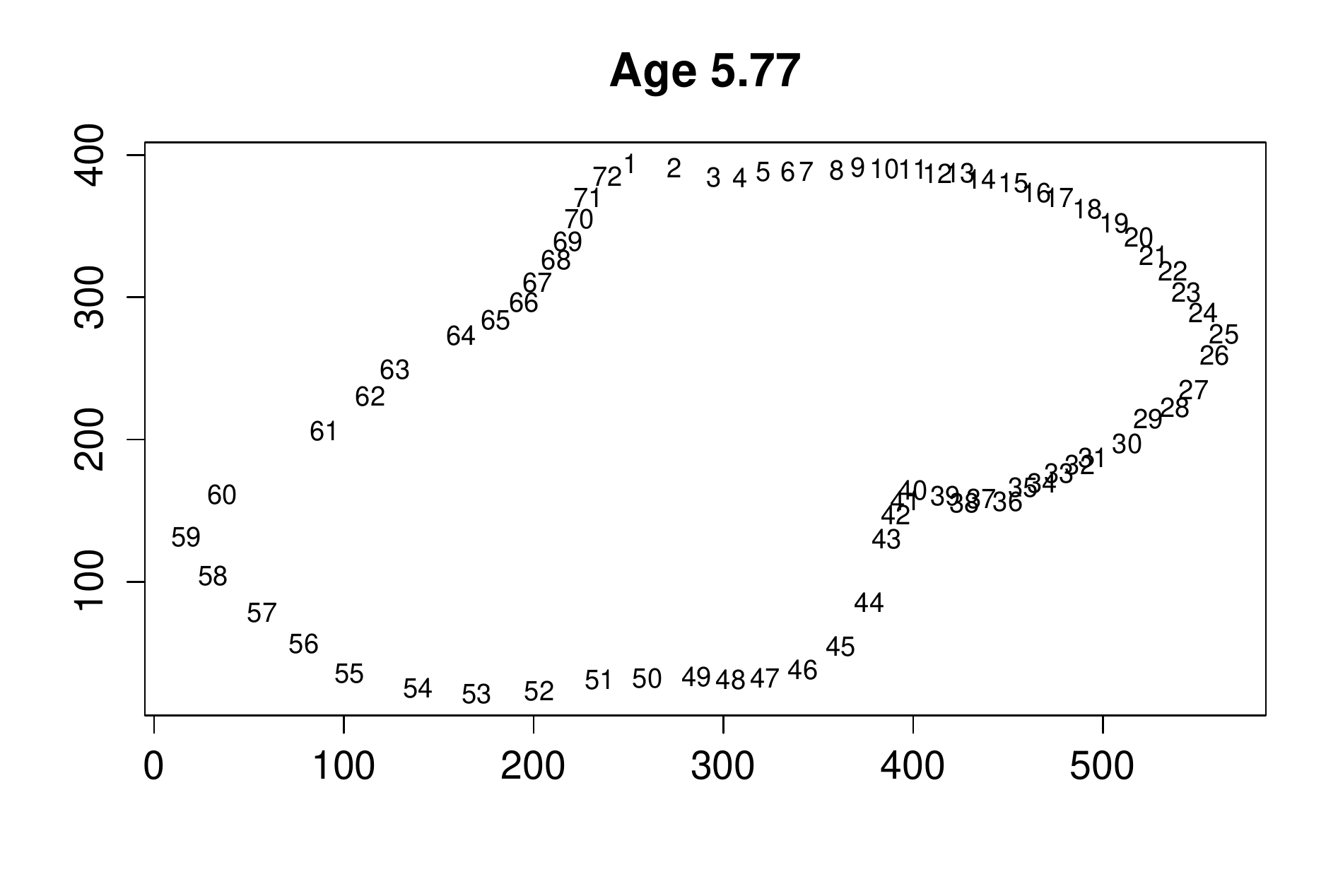}}
\caption{Macaca landmarks at 0.9 and 5.77 years.}
\label{fig:macaca}
\end{figure}

\begin{figure}[!hbt]
\centerline{\includegraphics[width=0.47\textwidth]{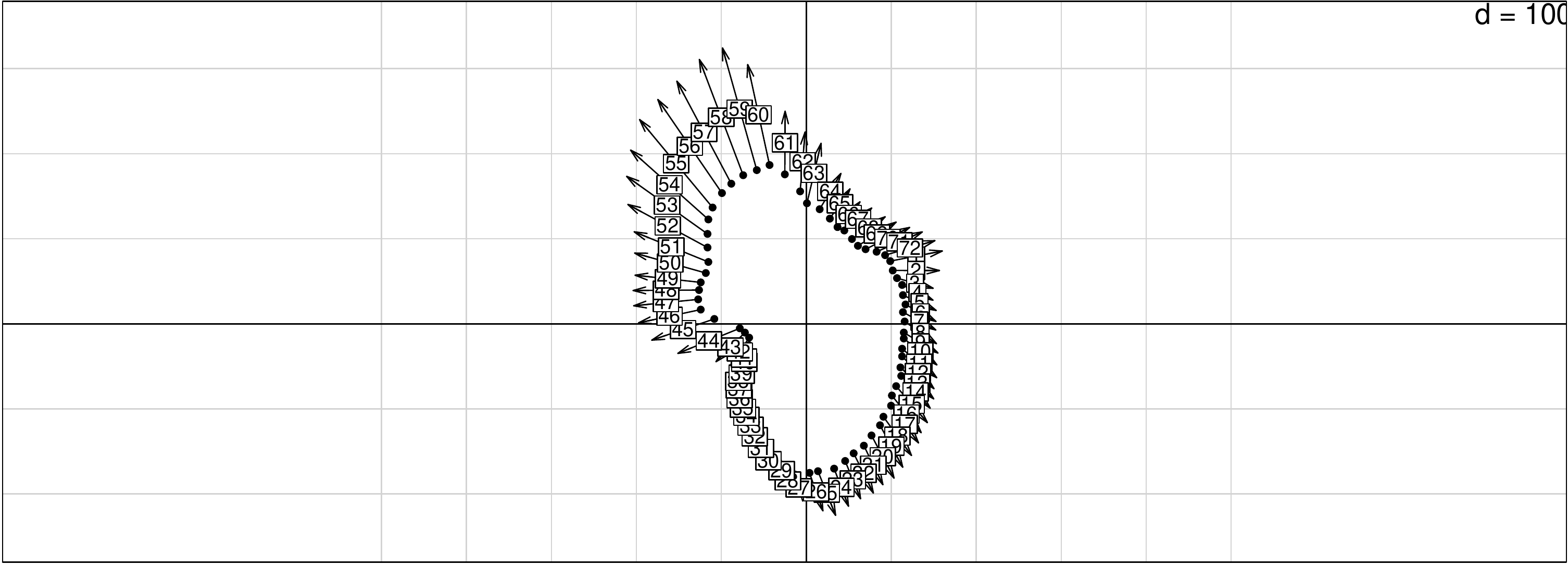} ~ \includegraphics[width=0.50\textwidth]{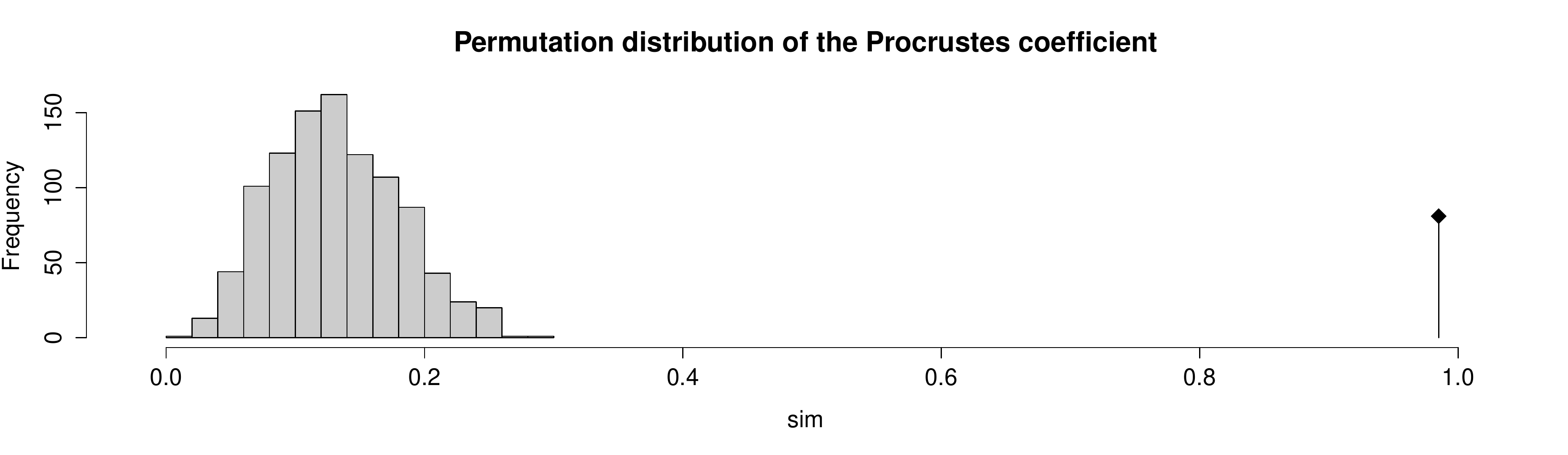}}
\caption{Left: Procrustes analysis to represent the deformation from 0.9 to 5.77 years of  the macaca face. Right: Permutation distribution of the Procrustes coefficient and its observed value. }
\label{fig:proc_permu}
\end{figure}

In cephalofacial growth studies, shape changes are analysed by collecting landmarks at different ages.
We focus here on a study on male Macaca nemestrina described in \citet{AJPA:AJPA1330590203}.
 Figure \ref{fig:macaca}  gives 72 landmarks of a macaca at the age of 0.9 and 5.77 years.
To study the similarity between the two configurations, we compute the association coefficients and their tests. 
The RV coefficient is 0.969 (its unbiased version is 0.94) and the dCor coefficient is 0.99 (its unbiased version is 0.985) and they are highly significant. 
The standard coefficient used on morphological landmark  data is the Procrustes coefficient described Section \ref{sec:proc}.
Procrustes analysis  superimposes different configurations as illustrated  Figure \ref{fig:proc_permu} on the left. The dots represent the shape at age 0.9 years and the arrows point to the shape at 5.77
years obtained after translation and rotation.  Figure \ref{fig:proc_permu} on the right represents the permutation distribution of the Procrustes coefficient under the null and the straight line indicates its observed value which is 0.984. The $p$-value associated to the test is thus very small.

\subsection{Chemometry data set}

In the framework of the EU TRACE project\footnote{\texttt{http://www.trace.eu.org}}, spectroscopic techniques are used to identify  and guarantee the authenticity of products such as the Trappist Rochefort 8 degree beer (one of seven authentic Trappist beers in the world). 
The data which were  presented as a challenge  at the annual French Chemometry meeting in 2010\footnote{ \texttt{http://www.chimiometrie.fr/chemom2010}} consist of 100 beers measured using three vibrational spectroscopic techniques: near infrared (NIR), mid-infrared (MIR) and Raman spectroscopy.
The beers were analysed twice using the same instruments, providing technical replicates. Table 5  shows the similarity  between the repetitions. Raman's spectral repetitions are stable whereas the other two methods are not.  
\begin{table}[ht]
\label{table:chemo_twice}
\begin{center}
\begin{tabular}{c c c c c}
  \hline
         & RV &RV$^{*}$ &dCor &dCor$^{*}$   \\  
NIR  &0.298 & 0.297&  0.709& 0.482 \\  
MIR  &0.597 &0.595 & 0.798 &0.585 \\  
Raman &0.978& 0.977& 0.987 &0.974 \\  
  \hline
\end{tabular}
\end{center}
\caption{Similarity between two measurements on the same 100 beers with different spectroscopic methods (NIR, MIR, Raman). RV coefficient and its debiased version RV$^{*}$  and the dCor coefficient and its debiased version dCor$^{*}$.} 
\end{table}
Table \ref{table:chemo_spectral} studies the similarities between measurments and shows that it provides complementary information since the values of the coefficient are quite small. 

\begin{table}[ht]
\label{table:chemo_spectral}
\begin{center}
\begin{tabular}{c| c c c ||  c c  c }
  \hline

& \multicolumn{3}{c}{RV$^*$ coefficient} &  \multicolumn{3}{c}{dCor$^*$ coefficient} \\
   \hline

   & NIR &  MIR&   Raman   &   NIR &  MIR&   Raman  \\  
NIR & 1 & 0.03 &      0.33     &      1 & 0.07 &    0.45          \\  
MIR  &           & 1& 0.03          &  & 1 & 0.05   \\  
Raman &        &&    1             &  & &1 \\  
   \hline
\end{tabular}
\end{center}
\caption{Similarity between the spectroscopic techniques (NIR, MIR, Raman). Debiased RV coefficient RV$^{*}$  and the debiased dCor coefficient dCor$^{*}$.} 
\end{table}


\section{Conclusion}

Technological advances now allows the
collection of many different types of data on samples (images, metabolic characteristics, 
genetic profiles or
clinical measurements). 
These heterogeneous sources of information can lead to improved explanatory resolutions
and power in the statistical analysis.
We have discussed  several coefficients of association  presented  
as functions of general dissimilarity (or similarity) matrices that are convenient for comparing heterogeneous data. 
We have outlined how to go beyond the calculation of these coefficients
to making sense of what causes the associations between these disparate sources of
information.

      
One strong point favoring the dCov test is that it is consistent against all dependent alternatives. On the other hand, the RV coefficient is designed to detect simple linear relationships. Although the use of relevant variable transformations can overcome this flaw such as 
the transformation of the  continuous variables into categorical factors. 

In practice, we recommend computation of  both the RV and the dCov coefficients and their debiased version
 to gain more insight into the nature of the relationships. 
In addition, we suggest  to supplement the study by a follow-up analysis with graphical methods to explore and visualize the complex multi-table data. We  described STATIS and MFA which rely on linear relationships between variables but the success with which these methods have allowed ecologists and food scientists to describe their data suggests that adapting them to incorporate nonlinear coefficients such as dCov could be a worthwhile enterprise. 

In this paper, we focused on the case of continuous variables and some comments can be made on the case of categorical variables or mixed variables both continuous and categorical. 
Users of multiple correspondence analyses \citep{Green06} have 
developed special weighting metrics for contingency tables and indicator matrices of dummy variables that
replace correlations and variances with chi-square based statistics. With these specific row and column weights, results are known for the RV coefficient: the RV coefficient between two groups of categorical variables is related to the sum of the $\Phi^2$ between all the variables and the RV between one group of continuous and one group of categorical variables to the sum of the squared correlation ratio $\eta^2$ between the variables \citep{Escoufier2006, Holmes2008, Pages14}. 

Results depend on the particular preprocessing choice (such as scaling), distance or kernel choices. This flexibility can be viewed as a strength, since many types of dependencies can be discovered. On the other hand, of course, it underscores the subjectivity of the analysis and the importance of educated decisions by the analyst.

\section*{Acknowledgements}

Julie Josse has received the support of the European Union, in the framework of the Marie-Curie
FP7 COFUND People Programme, through the award of an AgreenSkills fellowship (under grant
agreement n 267196) for an academic visit to Stanford.
Susan Holmes acknowledges support from the NIH grant R01 GM086884.

\bibliographystyle{plainnat}  
\bibliography{josse_j} 

\end{document}